\documentclass[
aps,%
11pt,%
final,%
notitlepage,%
oneside,%
twocolumn,%
nobibnotes,%
nofootinbib,%
superscriptaddress,%
noshowpacs,%
centertags]%
{revtex4}
\begin{document}
\selectlanguage{english}

\title{Spectral Variability of LBV star V\,532 (Romano's star)}

\author{\firstname{O.~N.}~\surname{Sholukhova}}
\affiliation{\saoname}

\author{\firstname{S.~N.}~\surname{Fabrika}}
\affiliation{\saoname}

\author{\firstname{A.V.}~\surname{Zharova}}
\affiliation{Sternberg State Astronomical Institute, Moscow
University, Moscow, 119992 Russia}

\author{\firstname{A.~F.}~\surname{Valeev}}
\affiliation{\saoname} \email{azamat@sao.ru}

\author{\firstname{V.P.}~\surname{Goranskij}}
\affiliation{Sternberg State Astronomical Institute, Moscow
University, Moscow, 119992 Russia}

\received{December 2, 2010}%
\revised{December 18, 2010}%

\begin{abstract}
We present the results of studying the spectral and photometric
variability of the luminous blue variable star V\,532 in M\,33.
The photometric variations are traced from 1960 to 2010, spectral
variations---from 1992 to 2009. The star has revealed an absolute
maximum of visual brightness \linebreak (1992--1994, high/cold
state) and an absolute minimum (2007--2008, low/hot state) with a
brightness difference of   $\Delta$B~$\approx 2.3^m$. The
temperature estimates in the absolute maximum and absolute minimum
were found to be $T \sim 22000$~K and $T \sim 42000$~K,
respectively. The variability of the spectrum of V\,532 is fully
consistent with the temperature variations in its photosphere,
while both permitted and forbidden lines are formed in an extended
stellar atmosphere. Broad components of the brightest lines were
found, the broadening of these components is
due to electron scattering in the wind parts closest to the
photosphere. We measured the wind velocity as a difference between
the emission and absorption peaks in the P\,Cyg type profiles. The
wind velocity clearly depends on the size of the stellar
photosphere or on the visual brightness, when brightness declines,
the wind velocity increases. In the absolute minimum a kinematic
profile of the V\,532 atmosphere was detected. The wind velocity
increases and its temperature declines with distance from the
star. In the low/hot state, the spectral type of the star
corresponds to WN8.5h, in the high/cold state---to WN11. We
studied the evolution of V\,532 along with the evolution of
AG\,Car and the massive WR binary HD\,5980 in SMC. During their
visual minima, all  the three stars perfectly fit with the WNL
star sequence by Crowther and Smith (1997). However, when visual
brightness increases, all the three stars form a separate
sequence. It is possible that this reflects a new property of LBV
stars, namely, in the high/cold states they do not pertain to the
{\it bona fide} WNL stars.
\end{abstract}

\maketitle

\section{INTRODUCTION}

Luminous blue variables (LBVs) represent one of the least
understood stages in the evolution of massive stars. Difficulties
in the understanding of these objects are dictated both by the
paucity of well-known LBV stars, their diversity, strong spectral
and photometric variability
\cite{Humphreys1994:Fabrika_n_en,vanGenderen2001:Fabrika_n_en}, as
well as by our inaccurate knowledge of their fundamental
parameters. Evidently, a massive star can manifest itself as an
LBV after the O star stage in the Main Sequence, and before the
late nitrogen WR star (WNL). However, a more accurate position of
LBV stars in the ${\rm OV} \to {\rm WNL}$ transition is yet
unclear \cite{Smith_Conti2008:Fabrika_n_en}. It is very likely
that the LBV stage and the wind instabilities, corresponding to
this stage are determined by the hydrogen abundance in the stellar
atmosphere. However, it is not an easy matter to both measure the
hydrogen abundance, and to precisely determine the luminosity and
temperature of the star. Measurement of fundamental parameters of
massive stars in nearby galaxies, such as the M\,33 galaxy, rich
in massive stars, may prove to be more confident than in our
Galaxy, since the distances to galaxies are measured with
sufficient accuracy.

LBV stars reveal strong spectral and brightness variations at a
roughly constant bolometric luminosity. Characteristic times of
such a global variability amount to months and years. When visual
brightness of an LBV star increases, the temperature of the
photosphere drops to $9000-10000$\,K, the size of the stellar
photosphere increases, and the star enters its high/cold state.
From here on, when we mention the photosphere, we refer to the
``pseoudophotosphere'', i.e. the place in stellar wind, where the
observed continuum radiation is formed. With a decrease in visual
brightness the temperature notably increases up to 35000\,K or
higher
\cite{Humphreys1994:Fabrika_n_en,Szeifert1996:Fabrika_n_en,Clark2005:Fabrika_n_en},
the size of the star decreases, it enters its low/hot state. In
the hot state the LBV spectra are very similar to the spectra of
WNL stars. However, the bona fide WN7/8/9-type stars differ from
the LBVs in the low/hot state by a number of spectral features
\cite{Smith_Conti2008:Fabrika_n_en,Smith1994:Fabrika_n_en} and low
hydrogen abundance. The so-called transitional (or ``slash'')
stars of the Ofpe/WN9-type, isolated in a separate class by
Walborn  \cite{Walborn1977:Fabrika_n_en} are similar to LBV stars
in the low/hot state. Ofpe/WN9 stars are characterized by a mixed
spectrum with emission lines, typical of the Of and WNL stars.
Taking an example of two galactic LBV-type stars AG\,Car and
He~3--519, which in the low/hot states correspond the
Ofpe/WN9-type, Smith, Crowther \& Prinja
\cite{Smith1994:Fabrika_n_en} proposed the extension of the WNL
classification up to WN10 and WN11.

Nevertheless, the question whether the bona fide LBV stars in the
low/hot states differ from the latest WN9/10/11 stars is still
open. In other words, is it possible that all the latest WNL-type
stars are in fact LBVs, which did not show any strong brightness
variability  over the time of their observations (only a few
dozens of years). Such stars, which may be suspected to belong to
the LBV type, are commonly called dormant LBVs. Note that dormant
LBV stars can be found not only among the WNL stars. Two classical
LBV  stars,  $\eta$\,Car and P\,Cyg now have a spectrum of
OB-hypergiants, their brightness variability is low. However, a
few hundred years ago they had giant eruptions
\cite{Humphreys1994:Fabrika_n_en}. If it were not for these giant
brightenings of $\eta$\,Car and P\,Cyg and the famous nebula
around $\eta$\,Car, we would not be able to suspect them in
belonging to the LBV class.

There are two LBV objects known that revealed \mbox{${\rm LBV}
\leftrightarrow {\rm WNL}$}-type transitions in recent years. This
is an LBV star AG\,Car in our Galaxy (see, for example,
\cite{Groh2009:Fabrika_n_en}) and an LBV star V\,532 in M\,33, to
which we devote this paper. A reverse transition is as well known,
\mbox{${\rm WN3} \to {\rm WN11(LBV)} \to {\rm WN4/5}$}, this is a
massive WR binary HD\,5980 in SMC (e.g.,
\cite{Koenigsberger2010:Fabrika_n_en}). The study of such stars is
particularly important. Obviously, the parameters of these stars
can be determined more accurately, since during the drastic
variations in their spectra neither the chemical composition, nor
the value of interstellar absorption (in some approximation, the
bolometric luminosity as well) would vary. The study of LBV stars
during such transitions might give us a chance to conclude about
the relation between the {\it bona fide} LBVs and WNL stars.

In this paper, we study the star V\,532 in the  M\,33 galaxy from
the spectra obtained from 1992 to 2009. During this period it
revealed its highest/cold (1992) and lowest/hot (1997--1998)
states, registered from the observations carried out since 1960.
The visual brightness difference between these two extreme states
amounted to $2.3^m$.

The star V\,532 (\cite{Artyukhina1995:Fabrika_n_en}\footnote[1]{in
the catalog its designation is M33 V0532, GCVS II/205, {\tt
http://cdsarc.u-strasbg.fr/viz-bin/Cat?II/205}; {\tt
http://cdsarc.u-strasbg.fr/viz-bin/Cat?B/gcvs} (GCVS B/gcvs).},
other names---Romano's Star, GR\,290) was discovered by G.~Romano
in \mbox{1978 \cite{Romano1978:Fabrika_n_en}.} He assumed that
this blue star belongs to the Hubble-Sandage-type variables.
However, without spectral data no definite conclusions could be
done. Furthermore, this star was studied photometrically in
\cite{Sharov1990:Fabrika_n_en, Kurtev2000:Fabrika_n_en,
Kurtev2001:Fabrika_n_en, Sholukhova2002:Fabrika_n_en,
Zharova2011:Fabrika_n_en}. Photometric behavior of this star is
quite typical for LBVs, it is very similar to the observed
brightness variation in  AG\,Car \cite{Stahl2001:Fabrika_n_en}.
Even the total amplitude of visual brightness variation of these
two stars is equal, it amounts to $2.3-2.5^m$.

The first spectrum of V\,532 was apparently obtained by
T.~Szeifert in 1992, and published in \mbox{1996
\cite{Szeifert1996:Fabrika_n_en}.} In his paper, this star was
classified as an LBV candidate. The second spectrum was obtained
in \mbox{1994 \cite{Sholukhova1997:Fabrika_n_en}} in a spectral
survey of LBV candidate stars in M\,33. The spectrum did not have
a very good quality, however, the authors suggested that this is a
WN-type star. Until 2000, V\,532 remained an LBV candidate, in a
well-known survey by Humphreys \cite{Humphreys1994:Fabrika_n_en}
it is also described as a candidate. Explicit spectral variability
of V\,532 was found from the spectra, obtained from 1992 to
\mbox{1999 \cite{Fabrika2000:Fabrika_n_en}.} Photometric
variability, followed by spectral variability, did not leave any
doubt that V\,532 is an LBV \cite{Fabrika2000:Fabrika_n_en}.

Later, V\,532 has been spectrally studied as an LBV star in
\cite{Polcaro2003:Fabrika_n_en, Fabrika2005:Fabrika_n_en,
Viotti2006:Fabrika_n_en, Viotti2007:Fabrika_n_en,
Valeev2009:Fabrika_n_en, Polcaro2010:Fabrika_n_en}, where new
evidences of spectral variability were  presented. From the
1998-2001  spectra,  when the star had an intermediate brightness,
Fabrika et al. \cite{Fabrika2005:Fabrika_n_en} found that its
spectrum corresponds to the Ofpe/WN9-type. Using quantitative
spectral criteria introduced \mbox{in
\cite{Crowther1997:Fabrika_n_en},} they classified the star as
WN10-11. In the same paper  \cite{Fabrika2005:Fabrika_n_en} a
nebula of low surface brightness was found around V\,532. The
stellar parameters were found in \cite{Valeev2009:Fabrika_n_en}
from the spectral energy distribution (SED), constructed based on
the optical \mbox{observations \cite{Massey2006:Fabrika_n_en},}
performed between autumn 2000 and autumn 2001, when the star was
in its intermediate brightness state (spectral class \mbox{WN10-11
\cite{Fabrika2005:Fabrika_n_en}}). The temperature of its
photosphere was estimated in \cite{Valeev2009:Fabrika_n_en} as $T
\sim 35000$~K, luminosity \mbox{$\lg(L/L\sun) = 6.24$,} and the
value of interstellar absorption $A_V \sim 0.8$. Since 2006 the
apparent brightness of V\,532 has heavily dropped, hence, the
temperature of the photosphere has increased. In this brightness
minimum the spectrum of the star was classified as WN8-9
\cite{Polcaro2010:Fabrika_n_en}.

In this paper, we present the longest series of spectral and
photometric observations of V\,532, based on which we trace the
variations in the stellar spectrum.


\section{OBSERVATIONS}

\subsection{Photometric Observations}

For the photometric estimates of the V\,532  we used the initial
data by Romano \cite{Romano1978:Fabrika_n_en}. These brightness
estimates were obtained based on 104 photographic plates by the
67-cm Schmidt telescope of the Asiago Observatory in 1960--1977.
We as well used the photo materials from the Sternberg
Astronomical Institute (SAI) of Moscow State University,
represented by 678 records, obtained on the 50-cm Maksutov
telescope (AZT-5) in 1973--2005. Photographic observations were
carried out in  a system, similar to the Johnson B filter. The SAI
photographic materials and the method of data reduction are
described in  \cite{Sharov2000:Fabrika_n_en,
Sholukhova2002:Fabrika_n_en}. All the photographic estimates were
reduced to the B system. The accuracy of these estimates is
approximately $0.1^m$, but it decreases and the error reaches
$0.2^m$ when the star gets weaker, near a plate edge the error can
reach $0.5^m$. A more detailed description of all the photographic
data is presented in \cite{Zharova2011:Fabrika_n_en}.

In 2001 we started the CCD observations of V\,532. They were
carried out on the 1-m Zeiss telescope of the Special
Astrophysical Observatory (SAO) and the two telescopes of the
Crimean Observatory of the SAI MSU---AZT-5 and 60-cm Zeiss-2
telescope. Errors of the CCD observations amount to $0.01-0.05^m$,
they are not worse than $0.05^m$. New BVRc standards were obtained
at the 1-m Zeiss telescope of the SAO RAS. They were also used for
the photographic \mbox{observations
\cite{Zharova2011:Fabrika_n_en}.} Hence, we finally have 104
estimates by \mbox{Romano \cite{Romano1978:Fabrika_n_en},} 645
estimates on the \mbox{AZT-5,} and multicolor CCD BVRc photometry
(65~B, 79~V, 46~Rc) in 2001--2010. We also use several estimations
by Viotti et al \cite{Viotti2006:Fabrika_n_en}, and one by
\mbox{Humphreys \cite{Humphreys1980:Fabrika_n_en}}. They are well
consistent with \linebreak our data.

\subsection{Spectral Observations}

The spectra of V\,532 were obtained mainly with the 6-m BTA
telescope using the UAGS, \mbox{MPFS
\cite{Fabrika2005:Fabrika_n_en}}  and SCORPIO
\cite{AfMois05:Fabrika_n_en} instruments. We also use the spectrum
of V\,532, courtesy of T.~Szeifert
\cite{Szeifert1996:Fabrika_n_en}, obtained in October 1992 on the
3.5-m Calar Alto telescope with the TWIN spectrograph, and the
spectrum obtained at our request by A.~Knyazev on the 2.2-m Calar
Alto telescope with the CAFOS spectrograph  in June 1999. We also
present here the spectrum we obtained on the SUBARU telescope with
the FOCAS spectrograph. It has been taken within the SS\,433
observational program \cite{Kubota2010:Fabrika_n_en} to make a
comparison of spectra of these two objects. SS\,433 spectrum  (hot
wind of the supercritical accretion disk) is very similar to the
spectra of WN\,10-11-type \mbox{stars
\cite{Fabrika2004:Fabrika_n_en}}. The log of spectroscopic
observations is given in the Table, which also presents the
spectral ranges and resolution for each observation. All the
spectra were processed by standard techniques, used for the
corresponding spectrographs in the MIDAS and IDL environments.

\begin{table*}[ht]
\setcaptionmargin{0mm} \onelinecaptionstrue
\caption{ Log of spectral observations}
\bigskip
\begin{tabular}{c|c|c|c|c}
\hline
JD 2400000\,+ & Date    &   Instrument / telescope  & Range, \AA &     Resolution, \AA \\
\hline
%
48910 & 10.15.1992 &  TWIN/Calar Alto  & 4450-5000 & 2.2\\
48910 & 10.15.1992 &  TWIN/ Calar Alto & 5800-6800 & 2.4\\

49366 & 01.13.1994 &  MOFS/BTA     & 5500-7600 & 8\\

51075 & 09.18.1998 &  MPFS/BTA     & 4450-5850 & 4.5 \\

51222 & 02.12.1999 &  UAGS/BTA     & 3280-8000 & 8.3 \\

51346 & 06.16.1999 &   CAFOS 2.2   & 5170-7690 & 7.5\\
51347 & 06.17.1999 &   CAFOS 2.2   & 3050-6050 & 7.5\\

51372 & 07.12.1999 &  UAGS/BTA     & 4330-5570 & 5.6\\
51372 & 07.12.1999 &  UAGS/BTA     & 5500-6740 & 5.6\\

51373 & 07.13.1999 &  UAGS/BTA     & 3410-8100 & 18\\

51408 & 08.17.1999 &  UAGS/BTA     & 4400-6860 &  7.5\\

51573 & 01.29.2000 &  UAGS /BTA    & 4350-6750 & 8.7\\

51821 & 10.03.2000 &  UAGS/BTA     & 4400-6820 &  8.3\\

51933 & 01.23.2001 &  UAGS/BTA     & 4310-6740  & 8.3\\

52150 & 08.28.2001 &  UAGS/BTA     & 4300-6730  & 9.4\\

52553 & 10.05.2002  &  MPFS/BTA     & 4000-7000  & 6   \\
53322 & 11.13.2004 &  MPFS/BTA     & 4000-7000  & 6  \\
53388 & 01.17.2005 &  MPFS/BTA     & 4000-7000  & 6  \\
53408 & 02.06.2005  &  SCORPIO/BTA  & 3500-7200  & 10  \\
53614 & 08.30.2005 &  SCORPIO/BTA  & 3500-7200  & 10  \\
53683 & 11.08.2005 &  SCORPIO/BTA  & 3500-7200  & 10  \\
53952 & 08.03.2006 &  SCORPIO/BTA  & 4000-5700  & 5   \\

54323 & 08.10.2007 &  SCORPIO/BTA  & 4000-5700  & 5  \\
54379 & 10.05.2007  &  SCORPIO/BTA  & 4000-5700  & 5  \\
54382 & 10.08.2007 & FOCAS/SUBARU  & 3750-5250 & 1.1   \\
54475 & 01.08.2008  &  SCORPIO/BTA  & 5700-7500  & 5  \\
54477 & 01.10.2008 &  SCORPIO/BTA  & 4000-5700  & 5  \\

55097 & 08.18.2009 & SCORPIO/BTA   & 3900-5700 & 5 \\
55115 & 10.09.2009 & SCORPIO/BTA   & 3100-7300 & 10 \\

\hline
\end{tabular}
\end{table*}

\section{RESULTS}

\subsection{Photometric Variability}

\begin{figure*}[ht]
\setcaptionmargin{15mm}
\onelinecaptionsfalse
\includegraphics[scale=0.6,angle=-90,trim=0mm 0mm 0mm 0mm]{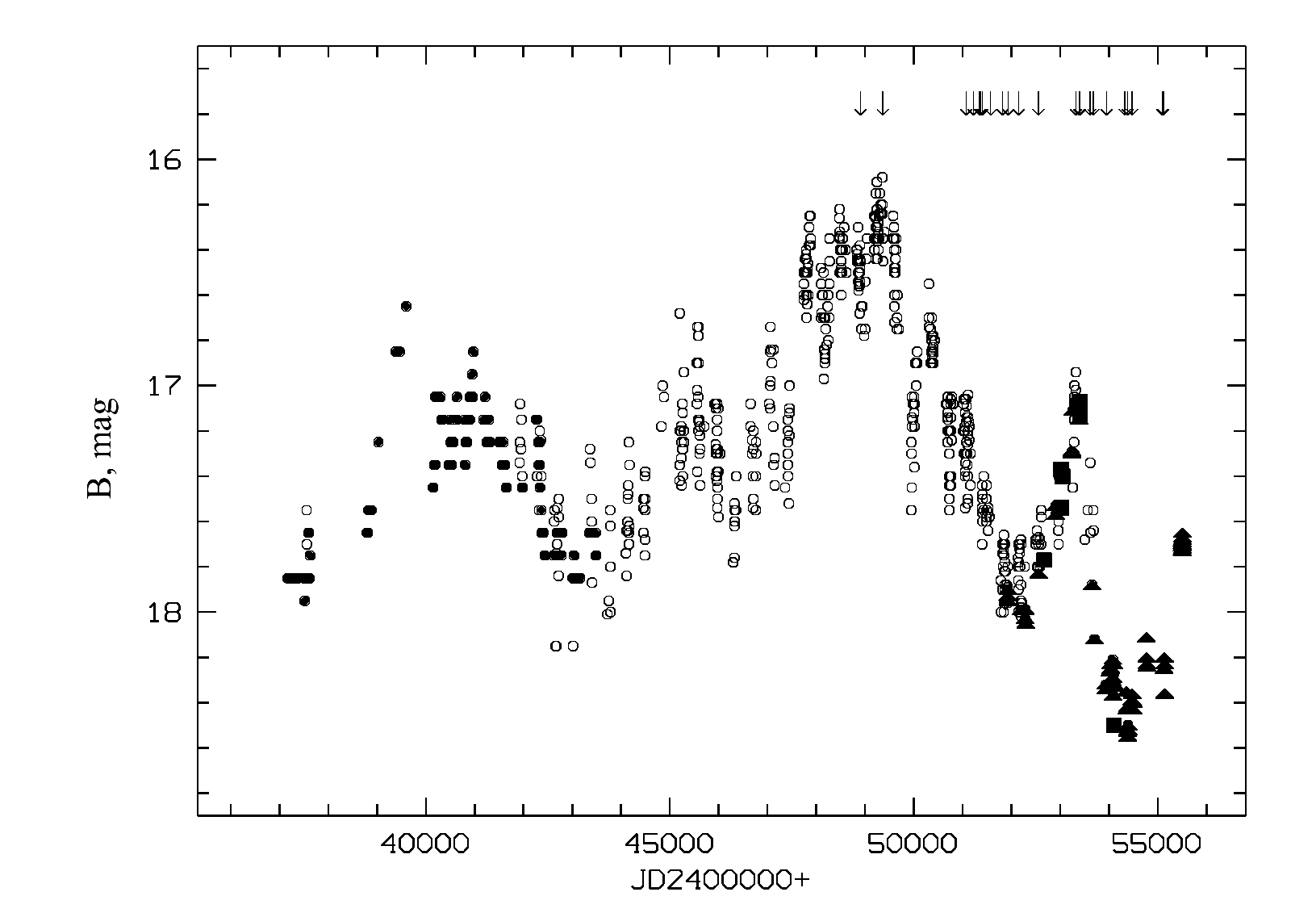}
\captionstyle{normal} \caption{Brightness variation of V\,532 in
the B-band based on all the data available from 1960 to 2010. Open
circles---photographic observations from
\cite{Zharova2011:Fabrika_n_en}, filled circles---photographic
observations by Romano \cite{Romano1978:Fabrika_n_en},
squares---data from \cite{Viotti2006:Fabrika_n_en},
triangles---CCD observations from \cite{Zharova2011:Fabrika_n_en}.
The arrows mark the timings of spectral observations. Over the 50
years of observations the star revealed an absolute maximum
between 1992--1994 (high/cold state) and an absolute minimum
between 2007--2008 (low/hot state).} \label{1:Fabrika_n_en}
\end{figure*}

Figure\,\ref{1:Fabrika_n_en} presents all the photometric
measurements of V\,532 from 1960 to 2010 in the B-band. The arrows
indicate the timings of the spectra. Over 50 years of observations
the star revealed four strong brightness increases, one of which,
occurring between 1990--1992, was the greatest. Consequently,
there were some brightness minima as well, the deepest of which
was in \mbox{2007--2008.} \linebreak The maximum
brightness difference amounted to $\Delta B \approx 2.3^m$.
Spectral observations cover both these extreme states of V\,532.
Such a brightness variability with an amplitude of $2.1^m$ in the
time interval of years is typical for S\,Dor-type  stars
\cite{vanGenderen2001:Fabrika_n_en}. The light curve of V\,532 is
very similar to that of \mbox{AG\,Car
~\cite{Groh2009:Fabrika_n_en},} which also revealed several
brightness extrema over the past 30 years with a total amplitude
of   $\Delta V \approx 2.4^m$ \cite{Groh2009:Fabrika_n_en,
Stahl2001:Fabrika_n_en}.

The scatter of the photometric estimates within one season of
observations is about $0.4^m$. It is determined not only by the
measurement errors, but by real brightness variability as well. In
the early photographic observations the error of one measurement
amounted to $0.1-0.2^m$, but later, in the photoelectric and CCD
observations it was several times lower, still, the scatter of
brightness estimates for  one season is as great, about
$0.2-0.3^m$. Sholukhova et al. \cite{Sholukhova2002:Fabrika_n_en}
studied the light curve of V\,532 in the B-band. A number of
quasi-periods were found with characteristic times of dozens of
days both in the high and low states. Some of them were quite
stable and were present in both states. The \mbox{authors
\cite{Sholukhova2002:Fabrika_n_en}} suggested that periods in the
range of $20-30$ days and shorter are linked with the stellar
pulsations, and longer periods are related to the  wind pulsations
driven by the wind instabilities. In hydrodynamic simulations of
nonlinear radial oscillations of hot LBV stars \mbox{Fadeev
\cite{Fadeyev2010:Fabrika_n_en}} has found that the periodic
brightness variations in the range of $6-31$ days are largely of
pulsational nature.

\begin{figure*}[t]
\setcaptionmargin{15mm}
\onelinecaptionsfalse
\includegraphics[scale=0.6,angle=-90,trim=0mm 0mm -10mm 0mm]{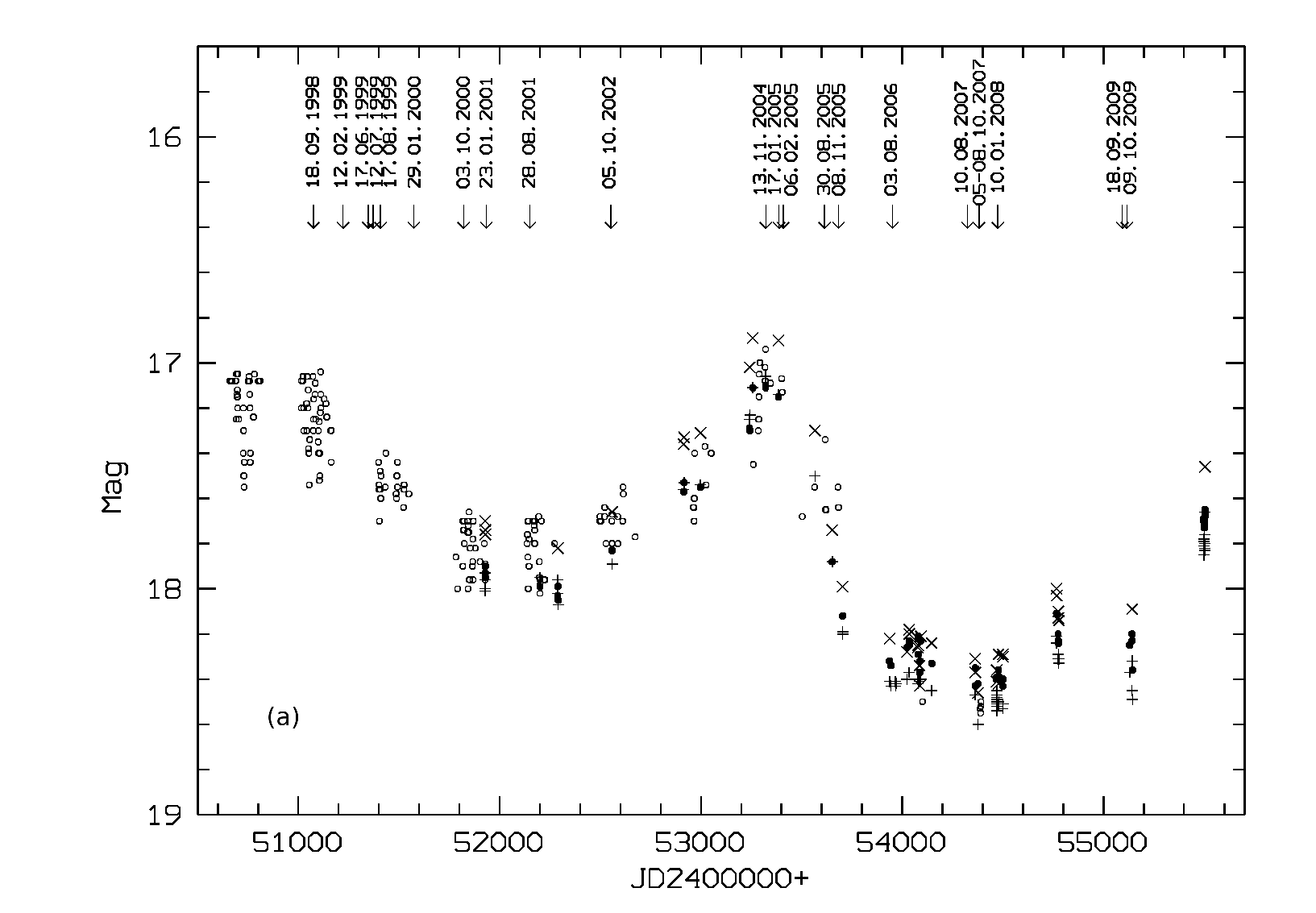}
\includegraphics[scale=0.6,angle=-90,trim=0mm -5mm 0mm 0mm]{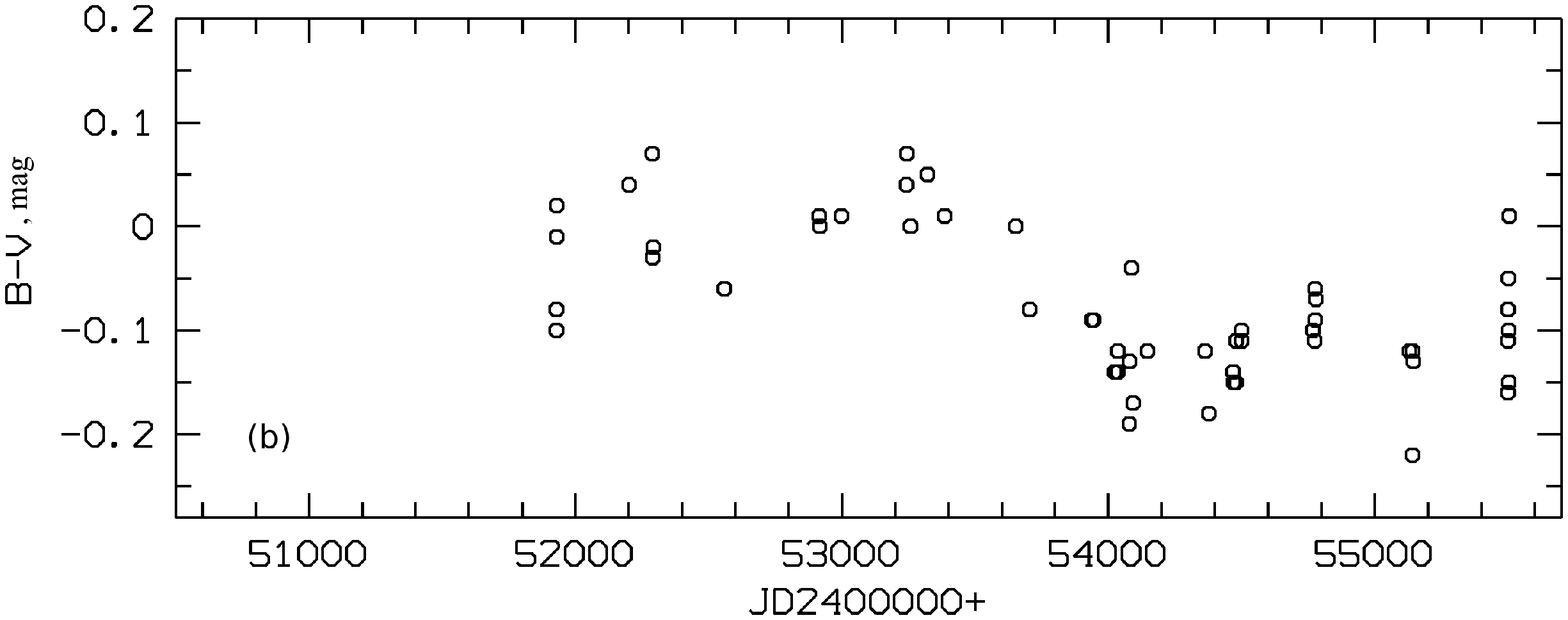}
\captionstyle{normal} \caption{(a): A fragment of the V\,532 light
curve, which is most densely covered by spectroscopic observations
(marked by arrows). Filled circles---B-band observations, straight
crosses---V-band, oblique \mbox{crosses---R-band,} open
circles---photographic observations. (b): variations of the B$-$V
color with time.} \label{2:Fabrika_n_en}
\end{figure*}

Figure\,\ref{2:Fabrika_n_en}a demonstrates a fragment of the light
curve of V\,532 in the B, V, R filters along with the photographic
measurements. Spectral observations cover the local minimum of
visual brightness observed in 2000--2002, the local maximum of
\mbox{2004--2005,} and the absolute minimum of \mbox{2007--2008}
quite well. In the late 2009 (from November) the brightness of the
star began to somewhat rise, and by November 2010 it increased by
$0.7^m$ in the \mbox{V-band.} Consequently, the minimum of
\mbox{2007--2008} was the deepest during the entire epoch of
observations of V\,532. The same figure shows the variation of
\mbox{B$-$V}
color during the epoch of
photoelectric and CCD observations. We can see that the color
variations are typical of LBVs: when the star's brightness
decreases, it grows bluer, however, the variations reveal some
irregularity.

\begin{figure*}[ht]
\setcaptionmargin{15mm}
\onelinecaptionsfalse
\includegraphics[scale=0.6,angle=-90,trim=0mm 0mm 0mm 0mm]{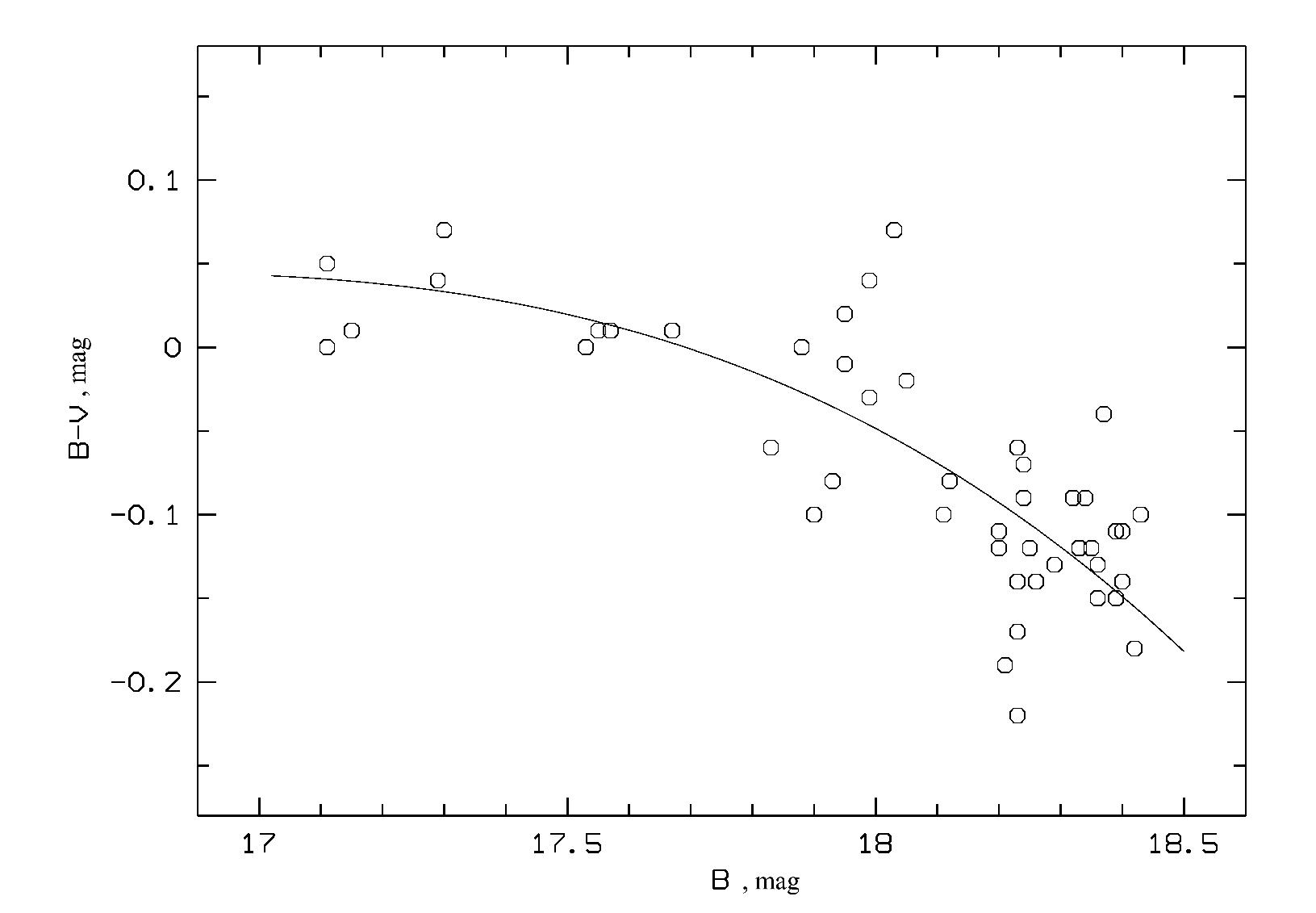}
\captionstyle{normal} \caption{B$-$V color variation with
brightness in the B-band. When the brightness decreases at
B~$>$17.9 the dependence gets steeper, and the amplitude of
irregular variations increases. The polynomial approximation is
shown (see text).
}
\label{3:Fabrika_n_en}
\end{figure*}

Figure\,\ref{3:Fabrika_n_en} demonstrates  the B$-$V color
dependence on the B-band brightness variations. In the low state
at B~$>$17.9 the color--magnitude dependency is very irregular,
the variation amplitude reaches $\Delta (B-V) \approx 0.15^m$. The
B brightness of about $17.9^m$ is isolated for another reason as
well, when the brightness is weaker, the color--magnitude diagram
gets steeper. If we describe this dependence by a simple function
without irregularities, we find: \small
$$
{\rm B}-{\rm V} = 0.037 + 0.033({\rm B} - 16.0)^2 - 0.027({\rm B}
- 16.0)^3.
$$
\normalsize \noindent We shall use this dependence further in the
analysis of line flux variability. We adopt the value of stellar
brightness in the deepest minimum as \mbox{B~$= 18.38^m$},
\mbox{V~$= 18.55^m$} with an uncertainty of about 0.05.

\subsection{Spectral Variability}
\begin{figure*}[htp]
\setcaptionmargin{5mm}
\onelinecaptionsfalse
\includegraphics[scale=0.6,angle=0,trim=10mm 0mm 0mm 0mm]{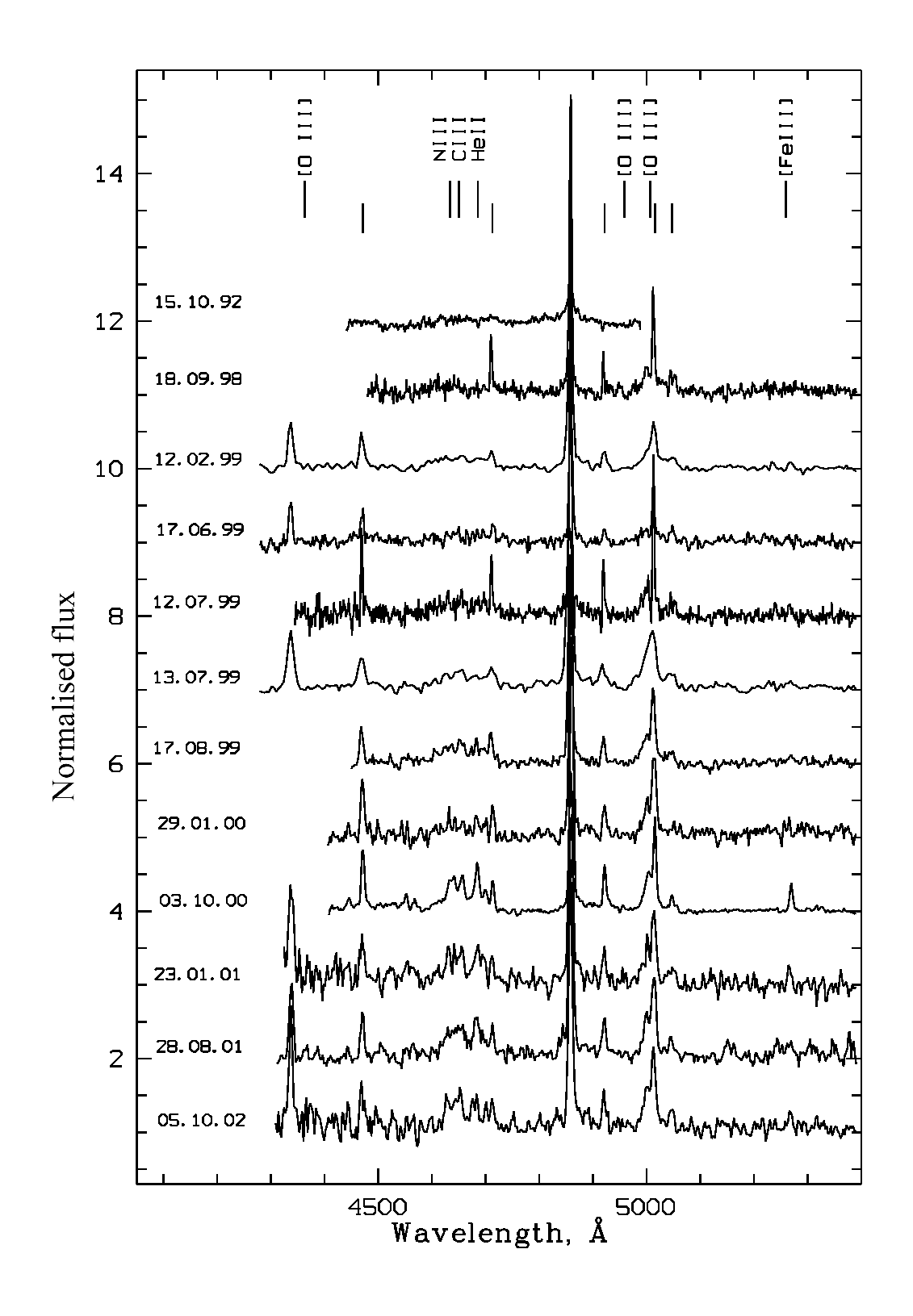}
\caption{Spectra of V\,532 in the blue region. The spectra are
normalized to the continuum and shifted along the vertical axis,
each by 1.0. Table~1 lists the spectral resolution for each
spectrum. Identifications of major lines that are either always
present, or appear in individual states of the star are shown. The
He\,I lines are marked by vertical segments without inscriptions.
The hydrogen lines are not marked not to overload the figure.}
\label{4:Fabrika_n_en}
\end{figure*}

\begin{figure*}[htp]
\setcaptionmargin{5mm} \onelinecaptionstrue
\includegraphics[scale=0.6,angle=0,trim=10mm 0mm 0mm 0mm]{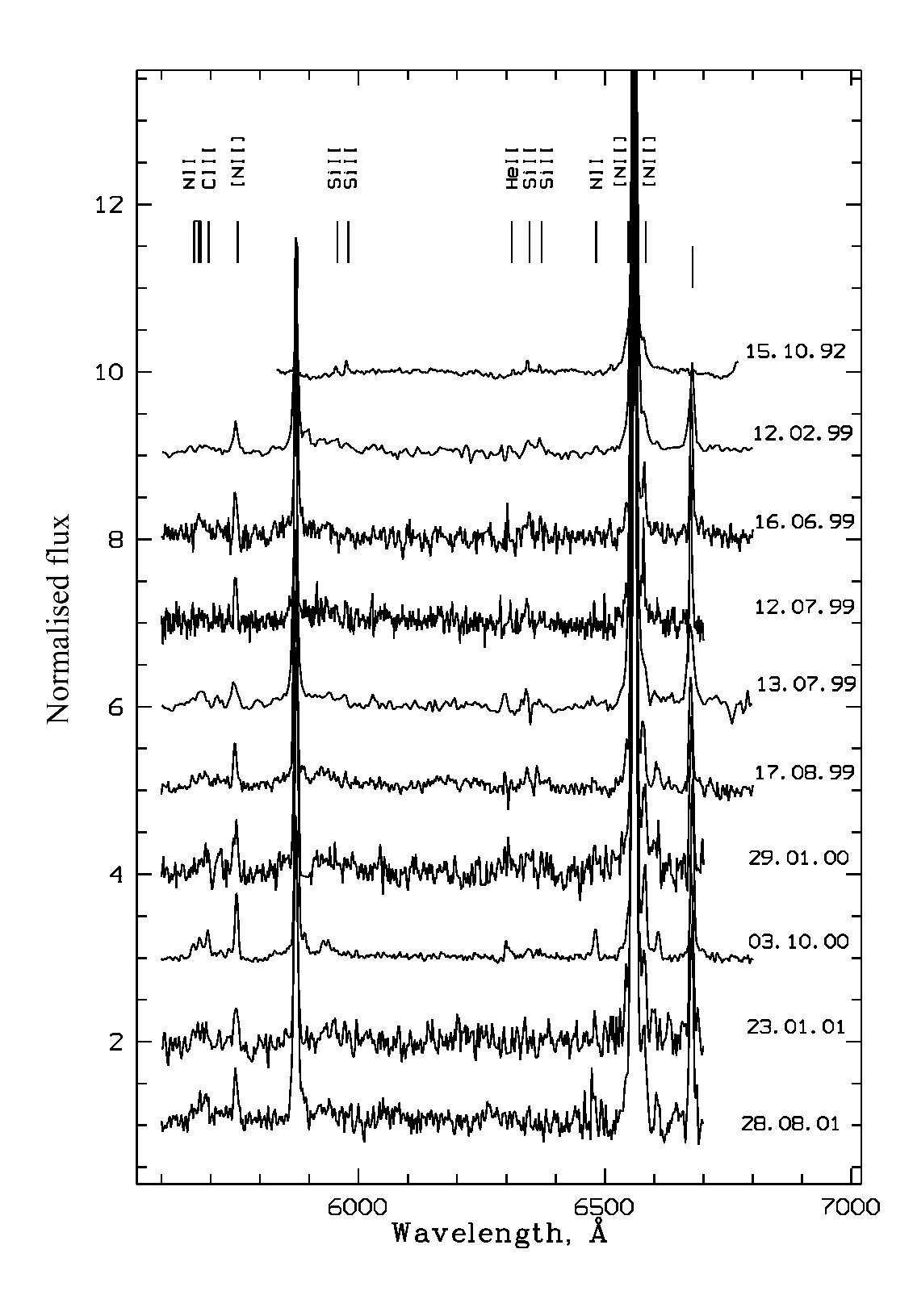}
\caption{Spectra of V\,532 in the red region, the rest as
specified in Fig.\,\ref{4:Fabrika_n_en}. } \label{5:Fabrika_n_en}
\end{figure*}

\begin{figure*}[htp]
\setcaptionmargin{5mm} \onelinecaptionstrue
\includegraphics[scale=0.6,angle=0,trim=10mm 0mm 0mm 0mm]{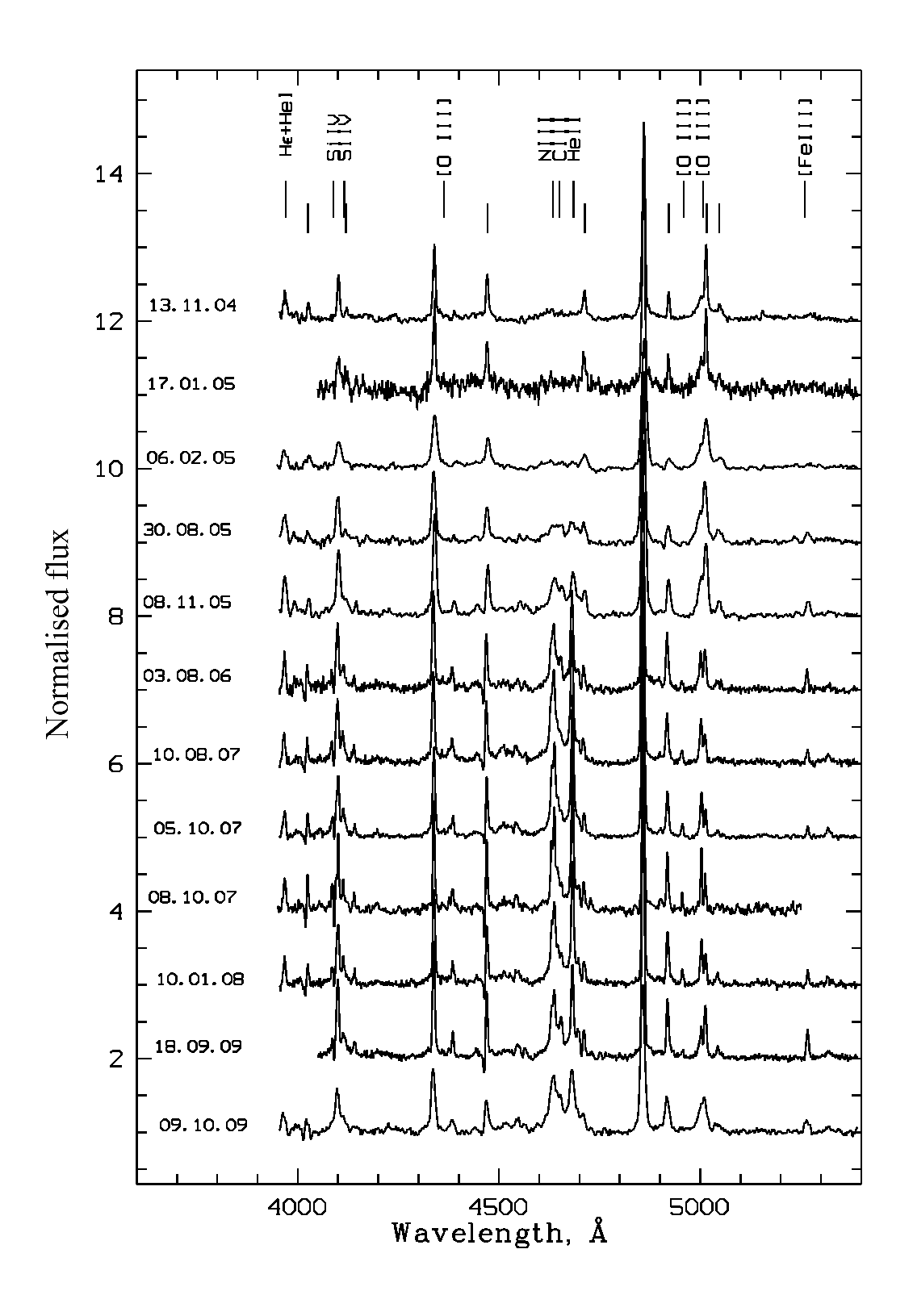}
\caption{Spectra of V\,532 in the blue region, the rest as
specified in Fig.\,\ref{4:Fabrika_n_en}. } \label{6:Fabrika_n_en}
\end{figure*}

\begin{figure*}[htp]
\setcaptionmargin{5mm} \onelinecaptionstrue
\includegraphics[scale=0.6,angle=0,trim=10mm 0mm 0mm 0mm]{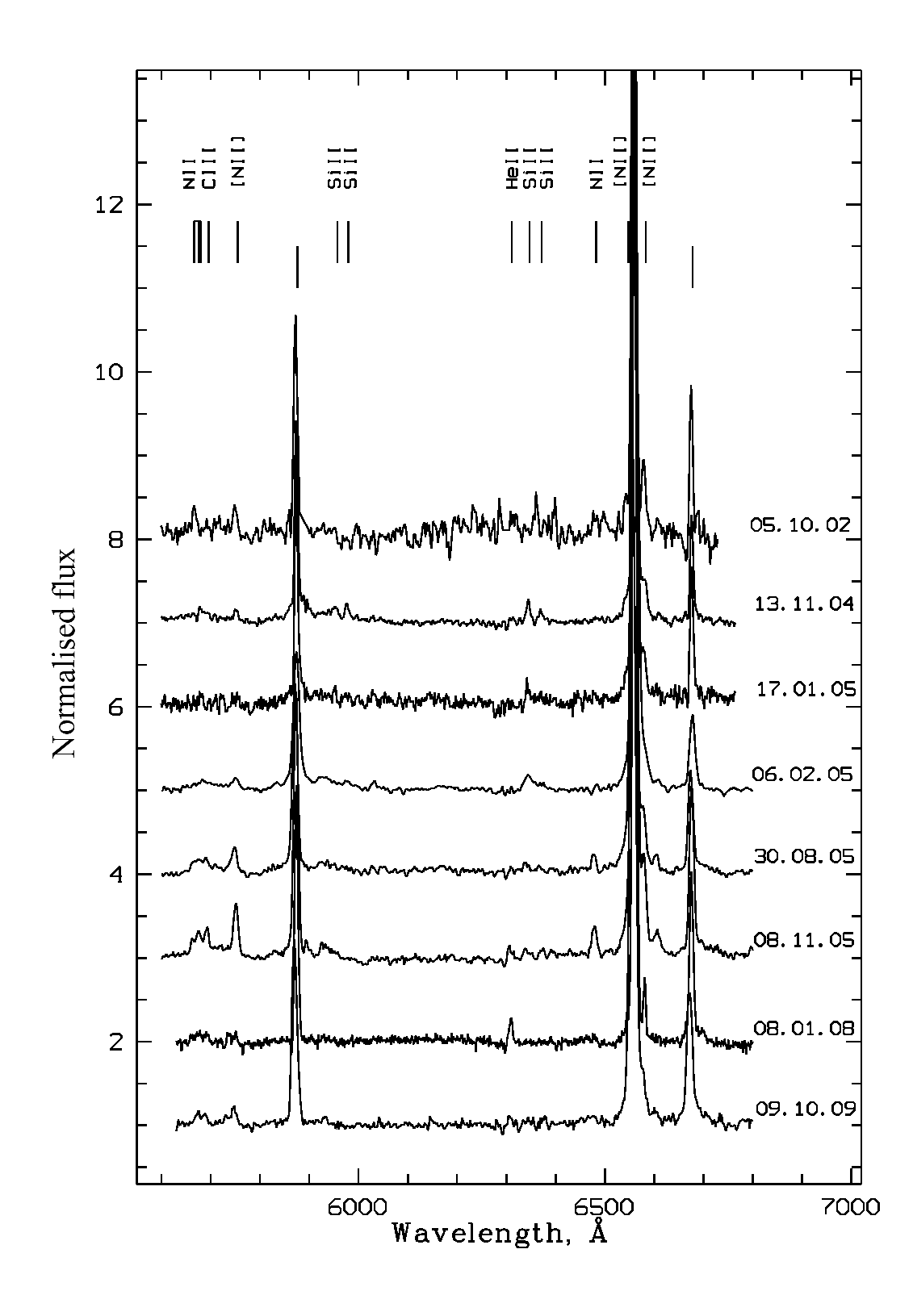}
\caption{Spectra of V\,532 in the red region, the rest as
specified in Fig.\,\ref{4:Fabrika_n_en}. } \label{7:Fabrika_n_en}
\end{figure*}

Spectral observations cover the period from 1992 to 2009.
Figures\,\ref{4:Fabrika_n_en}--\ref{7:Fabrika_n_en} demonstrate
all the spectra we have in  blue and red ranges. The
identifications of key lines are shown. However, we will not dwell
on the detailed identification of the spectrum of V\,532 in this
paper (see its description \mbox{in
\cite{Fabrika2005:Fabrika_n_en} and
\cite{Polcaro2010:Fabrika_n_en}}), rather, we will  describe here
the evolution of the spectrum and individual characteristic
spectral lines.

Significant variations are seen in HeI line intensities, the
appearance and disappearance of P\,Cyg type profiles, the
evolution of the HeII~$\lambda 4686$ line and the
CIII/NIII~$\lambda 4625-4650$ Bowen blend, the appearance of
forbidden lines [OIII]~$\lambda\lambda$4959, 5007, \linebreak
\mbox{[FeIII]~$\lambda\lambda$4658, 4702, 5270,}
[ArIII]~$\lambda$7135 (the latter is not shown in the figure) and
others.
The   HeII~$\lambda$5412~\AA\,line (not shown) appears
simultaneously with the strong HeII~$\lambda$4686 line, however,
it always has a P\,Cyg type profile. The latter is explained quite
well by a strong difference of optical depths in these lines, in
the winds of WN-stars the 5412~\AA\,line is formed close to the
photosphere ($2-3$ radii of the photosphere), while the 4686~\AA\,
line is formed at the distance of several tens of radii of the
\linebreak photosphere \cite{Hamann1995:Fabrika_n_en}.

When V\,532 goes through a transition from its high to low state,
the spectral variations in the star are quite obvious
(Fig.\,\ref{4:Fabrika_n_en}--\ref{7:Fabrika_n_en}), the
photosphere temperature increase is naturally followed by the
appearance and strengthening of high-excitation lines.
Spectroscopy covers two high states: one in \mbox{1990--1992} (the
absolute maximum) and the other in \mbox{2004--2005,} and two low
states: one in \mbox{2000--2002} and the other occurring in late
2006--early 2008 (the absolute minimum). Apart from obvious
variations in the spectrum, we would like to draw attention to the
evolution of the broad components of most of the brightest
hydrogen and helium lines. In the high state (1992 and
$2004-2005$) there appears a broad component in the H$\beta$ line,
which vanishes in the low state. In the intermediate high state
there appears a broad component in the He\,I~$\lambda$5876 line
(e.g., the 11.13.2004 spectrum), which is neither observed in the
higher or low states. Similar broad wings in the
He\,I~$\lambda$5876 line are found in AG\,Car
\cite{Stahl2001:Fabrika_n_en} exactly in the same intermediate or
close to the high brightness state. In the low state the broad
wings of this line disappear \cite{Groh2009:Fabrika_n_en}. When
V\,532 is in its low state, its spectrum reveals broad wings of
the HeII~$\lambda$4686 line, but they are not observed in the
HeII~$\lambda$5412 line. It follows from here that the broad
components appear in the lines with maximal optical depth and only
in the lines, for which the temperature of the photosphere is
optimal to excite this transition. Consequently, broad components
are in no way connected neither with the rotation of the star, nor
with the Doppler broadening of the wind lines, as there are bright
narrow components present in the same line profiles. The
appearance of broad line wings is linked with broadening owing to
the scattering of photons by free electrons in the most dense and
closest to the photosphere parts of the wind.

If the appearance of broad line wings is due to the scattering of
light by electrons, as usually happens with the hydrogen lines in
hot supergiants, the width of the line wings is determined by a
combination of optical depth in the given line and the electron
temperature. In V\,532, we measured the widths of the broad
components, and after correcting for spectral resolution we found
the following: FWHM(H$\beta$, H$\alpha) = 1100 \pm 20$~km/s
(spectrum of 10.15.92), FWHM(He\,I~$\lambda 5876) = 1800 \pm
100$~km/s (spectrum of 11.13.04). Other He\,I lines also reveal
large pedestals in the profiles, but they are difficult to
measure, FWHM(He\,II~$\lambda 4686) = 1500 \pm 70$~km/s (from the
spectra of $2007-2008$).

We measured the radial velocity of V\,532 from the spectrum,
obtained by T.~Szeifert on 10.15.92. From four Si\,II emission
lines, the profiles of which are completely symmetric
($\lambda\lambda 5958, 5979, 6347, 6371$) \linebreak the
heliocentric radial velocity amounts to \linebreak \mbox{$-183
\pm3$~km/s.} From the Si\,III~$\lambda 4553$ absorption line the
velocity is $-198$~km/s (absorption lines may well have a blue
shift due to the wind effects). From the forbidden line
[NII]~$\lambda 6583$ we obtained the velocity of $-180$~km/s. This
line most likely belongs to the extended \mbox{nebula
\cite{Fabrika2005:Fabrika_n_en},} which was captured by the
spectrograph slit and can not be fully subtracted. The hydrogen
lines are distorted by absorption from the blue side, the He\,I
lines are mainly in the absorption, they are very weak and may
quite be distorted by emission. Hence, we did not make any
velocity measurements from  these lines.


 We also measured radial velocities from the spectrum we
took on the Subaru telescope on \linebreak 10.08.2007. At this
time the star was in its low/hot state. This spectrum has good
quality, but almost all the major lines (He\,I) are distorted by
the absorption from the blue side. The peak of the symmetrical
H$\beta$ line revealed the velocity of  $-186$~km/s, two lines
[OIII]~$\lambda\lambda 4959\,$ and $5007$---the velocity of
$-220$~km/s. These [OIII] lines are  very variable, their
variation fully repeats the evolution of [FeIII]~$\lambda5270$,
[ArIII]~$\lambda$7135, [NII]~$\lambda$5755 and some permitted
lines (see below). We believe that all these forbidden lines are
formed in the outer extended stellar atmosphere, the asymmetry of
which may lead to a distortion of radial velocities. Focusing on
the good agreement between the radial velocities, measured from
symmetrical  Si\,II and H$\beta$ emission lines in the spectra,
obtained with the time difference of 15 years with a better
spectral resolution, we adopt the radial velocity of the star as
$-184 \pm3$~km/s.

\subsection{Spectral Classification}

In the intermediate brightness state in \mbox{$1998-2001$}
(Fig.\,\ref{2:Fabrika_n_en}), Fabrika et al.
\cite{Fabrika2005:Fabrika_n_en}, using quantitative spectral
criteria introduced by Crowther and Smith
\cite{Crowther1997:Fabrika_n_en} found that the spectral type of
V\,532 corresponds to \mbox{WN10-11.} Later, based on the 2008
spectra, when the star was in its lowest state, Polcaro et al.
\cite{Polcaro2003:Fabrika_n_en} using the same criteria
\cite{Crowther1997:Fabrika_n_en} found that the spectral class of
the star is close to WN8-9. It appears that depending on the value
of visual brightness the spectral class of V\,532 varies between
WN11 and WN8-9.

\begin{figure*}[ht]
\setcaptionmargin{15mm}
\onelinecaptionsfalse
\includegraphics[scale=0.45,angle=0,trim=0mm 0mm 0mm 0mm]{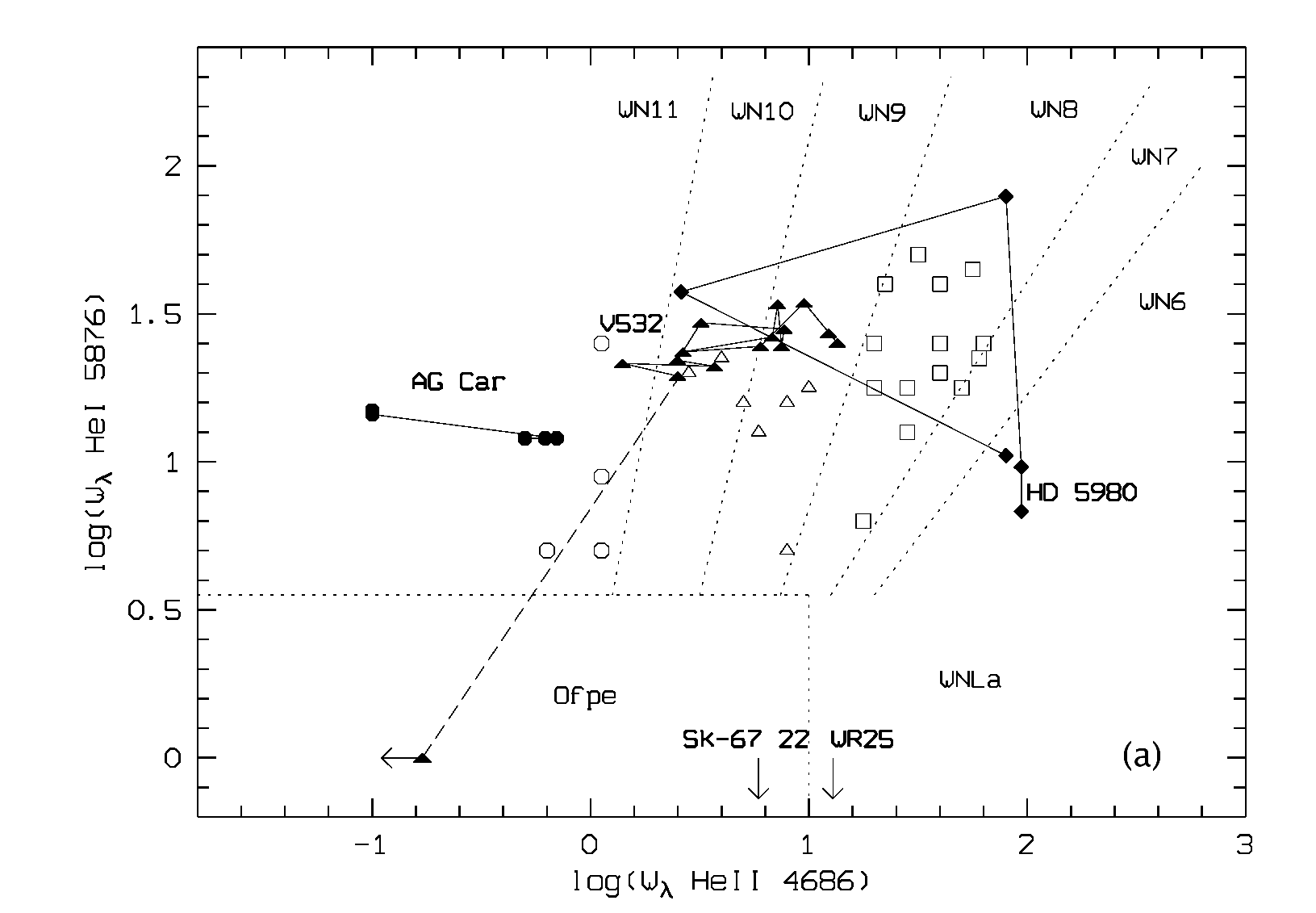}
\includegraphics[scale=0.45,angle=0,trim=0mm 0mm 0mm 0mm]{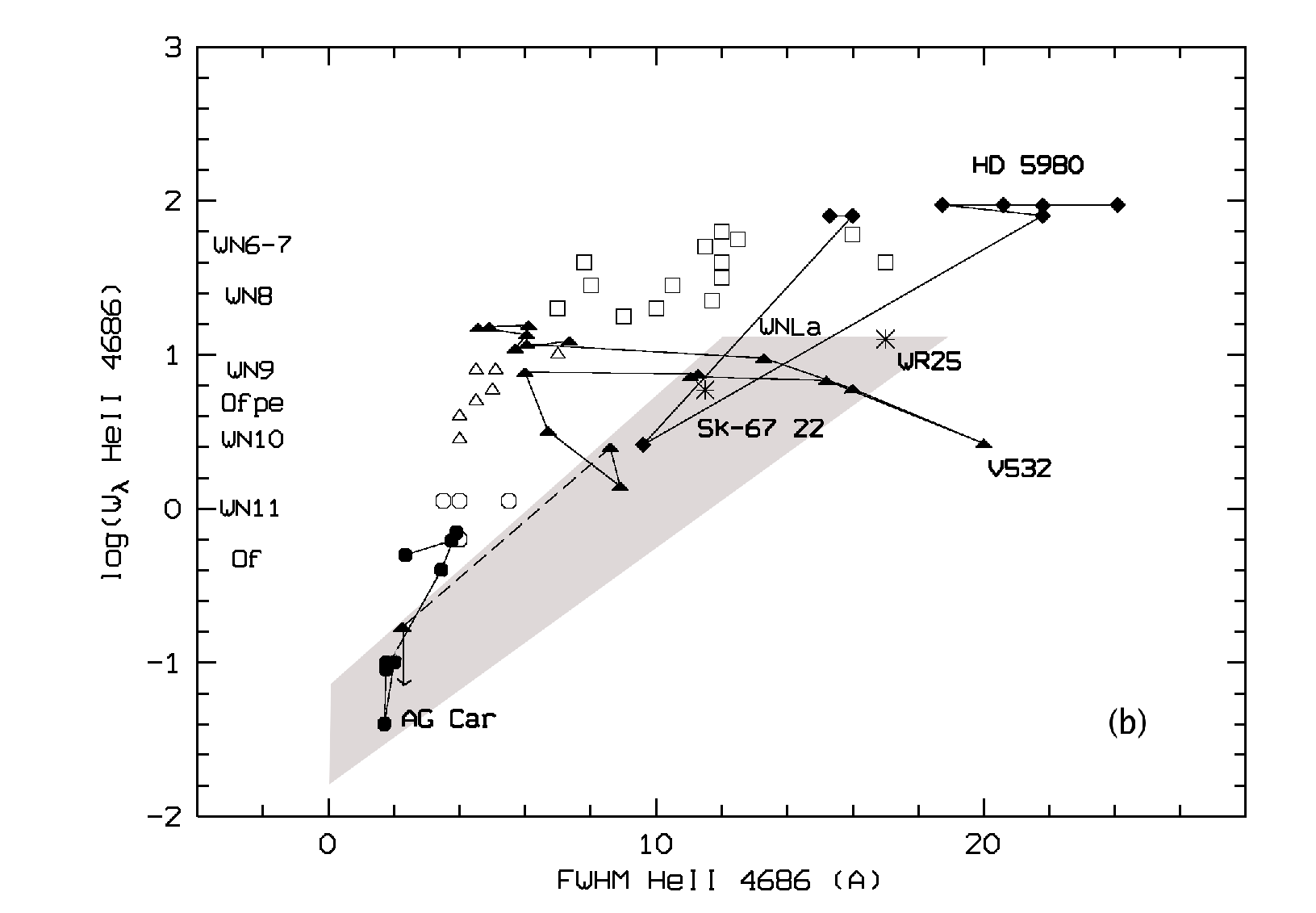}
\captionstyle{normal} \caption{The classification diagrams from
\cite{Crowther1997:Fabrika_n_en} for the WNL sequence. The open
signs mark the following stars: WN8 (squares), WN9--10
(triangles), and WN11 (circles), as well as  WR\,25 and
Sk~$-67^{\circ}~22$ (arrows and asterisks). The movements of
AG\,Car and HD\,5980 are also shown during their LBV--WNL
transitions. During the minima of visual brightness in the low
state all the three stars fit very well into the sequence of WNL
stars. When the brightness is increased, AG\,Car, V\,532 and
HD\,5980 form an individual sequence, which is not consistent with
the WNL sequence~~\cite{Crowther1997:Fabrika_n_en}, going beyond
its limits. On the bottom plot this new sequence is marked by a
gray background. } \label{8:Fabrika_n_en}
\end{figure*}

Having the longest series of spectroscopic observations we can
analyze the variability of V\,532 in more detail.
Figure\,\ref{8:Fabrika_n_en} reproduces the basic diagrams from
\cite{Crowther1997:Fabrika_n_en}, describing the classification
(mainly temperature)  sequences of late WN stars. In
\cite{Crowther1997:Fabrika_n_en} this diagram has been drawn up
for the stars of the Galaxy and LMC. However, the authors have
concluded that there is no fundamental difference between the
stars of these two galaxies in the classification diagrams. The
diagrams from Figs.\,\ref{8:Fabrika_n_en}a--\ref{8:Fabrika_n_en}b
show the relationship between the equivalent widths of
He\,I~$\lambda 5876$ and HeII~$\lambda$4686, and the relation
between the equivalent width and the width of the HeII line. The
widths of the HeII line (FWHM) in V\,532 were corrected for
instrumental resolution.

We plotted the stars WR\,25 and Sk~$-67^{\circ}~22$ \mbox{from
\cite{Crowther1997:Fabrika_n_en}} to Fig.\,\ref{8:Fabrika_n_en}.
In addition, we plotted there the stars AG\,Car
\cite{Groh2009:Fabrika_n_en,Stahl2001:Fabrika_n_en} and
HD\,5980~\mbox{\cite{Koenigsberger2010:Fabrika_n_en,
Barba1995:Fabrika_n_en, Heydari-Malayeri1997:Fabrika_n_en,
Foellmi2008:Fabrika_n_en}}, which are believed to have shown LBV
transitions (see Introduction). The movements of these two stars
in the diagrams reflect these transitions. For HD\,5980 we only
took the measurements that fall within the orbital phase of good
visibility of the A component of the
star~\cite{Koenigsberger2010:Fabrika_n_en}, which revealed the
WN\,11 spectrum during its famous outburst in October 1994. We can
see in Fig.\,\ref{8:Fabrika_n_en} that during this outbreak
HD\,5980 has indeed moved into the WN10--11 region in both
diagrams. The LBV event in HD\,5980 is clearly visible in the
figure. Until this outbreak (one point in Fig.~8a for 1993 and two
points in Fig.~8b for 1961 and 1993) the star was located in the
WN6-7 region. The point following the flare (December 1994) in no
way differs from all the subsequent WN6-7 state of the system
($1999-2005$) in Fig.~8b, but it differs by an increased
brightness of the He\,I, and He\,II lines (WN8) in Fig.~8a.

The movements of AG\,Car are not so radical, however, they fully
correspond to the modern ideas on LBV transitions. We marked in
Fig.\,\ref{8:Fabrika_n_en}a the transition from the absolute
minimum (1989, three points to the right) to the state of local
minimum (2001 and 2002, points to the left), according to
\cite{Groh2009:Fabrika_n_en}. In the diagram in
Fig.\,\ref{8:Fabrika_n_en}b~\cite{Groh2009:Fabrika_n_en,Stahl2001:Fabrika_n_en}
a group of topmost points marks the state of absolute minimum
(1989--early 1990), the lowest point---one of the spectra of the
absolute maximum (1995), the points in between mark an
intermediate state of 1991 and a local minimum \linebreak of 2001.

Hence, the movements of AG\,Car and HD\,5980 along the WNL
sequence are obvious. The star V\,532 at the states of increased
brightness fits well in the sequence between the low/cold state of
the LBV AG\,Car and the LBV episode of HD\,5980. In the low/hot
state all the three stars are perfectly consistent with the WNL
sequence~~\cite{Crowther1997:Fabrika_n_en}. However, in general, V\,532 moves
through the diagrams from Figs.\,\ref{8:Fabrika_n_en}a--\ref{8:Fabrika_n_en}b in a complex
manner.


In Figs.\,\ref{8:Fabrika_n_en}a--\ref{8:Fabrika_n_en}b the lowest
point of V\,532 (the spectrum from 10.15.1992 in the absolute
brightness maximum) indicates the upper limit of the possible
estimate of the equivalent width of HeII line, this line is
absent. In Fig.\,\ref{8:Fabrika_n_en}b we adopted the line width
equal to the spectral resolution in the spectrum. The
HeI~$\lambda$5876 line is confidently measured. Despite the
absence of HeII line in the spectrum of absolute maximum, the
upper limit of its equivalent width, as well as the HeI line
indicate that V\,532 could pass from the states with smaller
brightness to the state of absolute maximum through the region
WN11, in which the star AG\,Car is located. On the other hand, in
the absolute minimum ($2007-2008$ and 2009) in
Fig.\,\ref{8:Fabrika_n_en}a V\,532  is located in the far-right
position (region WN9), and in Fig.\,\ref{8:Fabrika_n_en}b---in the
upper left corner, exactly halfway between WN8 and WN9 (WN8.5).

\begin{figure}[t]
\setcaptionmargin{5mm} \onelinecaptionsfalse
\includegraphics[scale=0.45,angle=0,bb=80 35 550 705, clip]{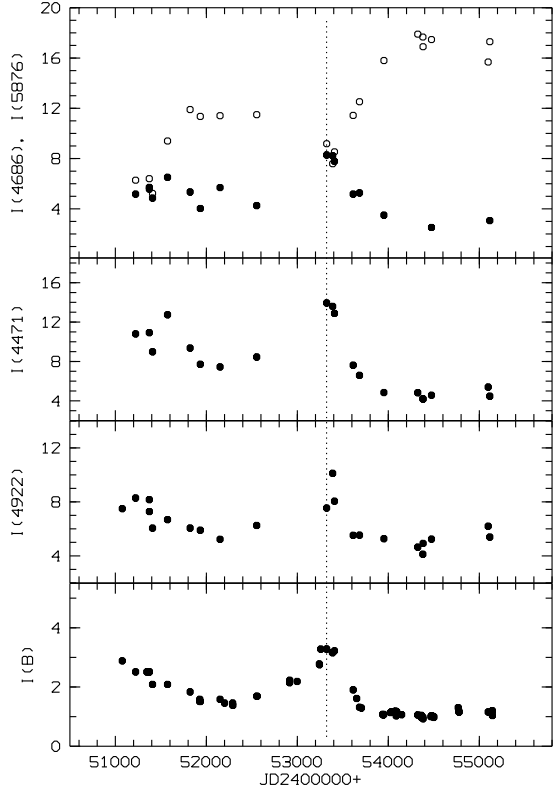}
\captionstyle{normal} \caption{Variations in relative intensities
of major HeI lines (filled circles) and
 He\,II~$\lambda 4686$ line (empty circles) from 1999 to 2009. These
relative intensities are the equivalent line widths, corrected for
the brightness variability in the corresponding filter, e.g.,
${\rm I} =$~EW~I(B)  (see text). The bottom plot shows the
relative intensity I(B), normalized to the B-band intensity during
the absolute minimum. The approximate position of the center of
local maximum of ~$2004-2005$ is marked by the vertical dotted
line.} \label{9:Fabrika_n_en}
\end{figure}

The evolution of the spectrum of V\,532 is as follows. Immediately
after the absolute maximum the star moves from the WN11 region
(the LBV region)  into the WN10-11 region in Fig.~8a, and also
into the WN10-11 region in Fig.~8b ($1998-1999$, the state before
the local minimum), this transition is marked by the dotted line
in \mbox{Figs.\,\ref{8:Fabrika_n_en}a--\ref{8:Fabrika_n_en}b}.
Further, in the local minimum of $2000-2002$, the star moves to
the WN9 region towards its position in the absolute minimum.
Later, the behavior of V\,532 becomes unusual in the transition to
the local maximum of $2004-2005$. At this time HeII line
distinctly broadens. In Fig.\,\ref{8:Fabrika_n_en}b the star
passes the region between two known objects WR\,25 (HD\,93162
WN6ha) and \mbox{Sk~$-67^{\circ}~22$} (O3\,If/WN6), it leaves the
main sequence of WNL stars, and its spectral class (formally)
becomes WN9--10, in Fig.~8a in this state the star becomes WN10.

The stars WR\,25 and Sk~$-67^{\circ}~22$ possess very high
luminosities and high hydrogen abundances in the atmospheres
\cite{Crowther1997:Fabrika_n_en}. The temperatures of these stars
are high enough, HeI lines are practically absent in their spectra
(Fig\,\ref{8:Fabrika_n_en}a and \cite{Crowther1997:Fabrika_n_en}).
The movement of V\,532 in the direction of WR\,25 and
Sk~$-67^{\circ}~22$ in Fig.\,\ref{8:Fabrika_n_en}b is not related
to the decrease in the size of the photosphere and the
corresponding increase of the escape velocity of the star (i.e.,
the wind velocity, as it likely occurs in the stars WR\,25, and
Sk~$-67^{\circ}~22$), but rather with the HeII line broadening as
a result of Thomson scattering.

By contrast, in  2004--2005 the size of the photosphere of V\,532
increased, as the stellar brightness grew in the visual region.
Following simple logic, He\,II line should become narrower, since
the escape velocity of the star has dropped, but in fact it
broadened. This broadening has caused V\,532 to move away from the
general sequence. We associate the He\,II line broadening with the
scattering of photons of this line, by the electrons. The most
extreme point in this figure with \mbox{FWHM(He\,II) $\sim
20$\,\AA\,} is uncertain, since the line has noticeably weakened
at the time. In addition, the He\,II~$\lambda 4686$ line is
located in a complex blend. Below, we describe the two-component
profile of the He\,II line, however, in
Figs.\,\ref{8:Fabrika_n_en}a--\ref{8:Fabrika_n_en}b all the width
measurements were made under the assumption of a single-component
profile. The remaining measurements of the He\,II line  width in
this figure are quite reliable.

When the visual brightness increases, all the three stars,
AG\,Car, V\,532 and HD\,5980 form a separate sequence, which is
not consistent with the WNL sequence
\cite{Crowther1997:Fabrika_n_en} and goes beyond its limits. In
times of visual brightness minima all the three stars perfectly
fit into the sequence of WNL stars. This result is important.
Perhaps the departure from the WNL sequence is related to the HeII
line broadening due to the same Thomson scattering with increasing
optical depth in this line. However, it is possible that this is a
new property of LBV stars, in the high state they do not
correspond to the latest {\it bona fide} WNL stars. We know only
three examples of such transitions, and naturally, additional
observations and new objects are required for more reliable
conclusions.

From the spectral classification of WN
stars~\cite{SmithMoffat1996:Fabrika_n_en} by several criteria we
additionally found the spectral class of V\,532. The ratio of
\mbox{HeII~$\lambda 5412$/HeI $\lambda 5876$} line  intensities,
defined as the  peak/continuum ratio (the first line has a clear
P\,Cyg type profile, hence the ratio of equivalent widths gives an
inaccurate result and a large scatter) appears to be equal to
$0.26 \pm 0.02$, which means the WN8 type (possibly, WN8.5). It is
interesting that this ratio is equal both in the absolute minimum,
and in the local maximum of $2004-2005$. In the spectra of the
absolute minimum the ratio of lines N\,V$\lambda
4604$/N\,III$\lambda 4640 \la 0.05-0.08$, what also indicates
spectral type \mbox{WN8-9 \cite{SmithMoffat1996:Fabrika_n_en}}.

The ratio N\,IV~$\lambda 4058$/N\,III~$\lambda 4640 \approx 0.14$
in the state of absolute minimum (specifically, from the Subaru
spectrum of 10.08.2007) also yields \mbox{WN8-9.} The line
CIV~$\lambda 5808$  is not found in this minimum, it also suggests
that the star has the WN8 class  or later. The relative abundance
H$^+$/He$^{++}$, calculated by the ratio of lines (H~$+$~He) both
from \mbox{4861 \AA{}}, and from \mbox{4340 \AA{}} to the
geometric mean of their neighboring Pickering lines
~~\cite{SmithMoffat1996:Fabrika_n_en} appears to be equal to
\mbox{$18 \pm 6$}, which means a subclass ``h'', i.e. an explicit
presence of hydrogen.

\subsection{Temperature Variations in the Photosphere \\ and Wind}

\begin{figure}[ht]
\setcaptionmargin{10mm} \onelinecaptionstrue
\includegraphics[scale=0.47,angle=0,trim=0mm 0mm 0mm 0mm]{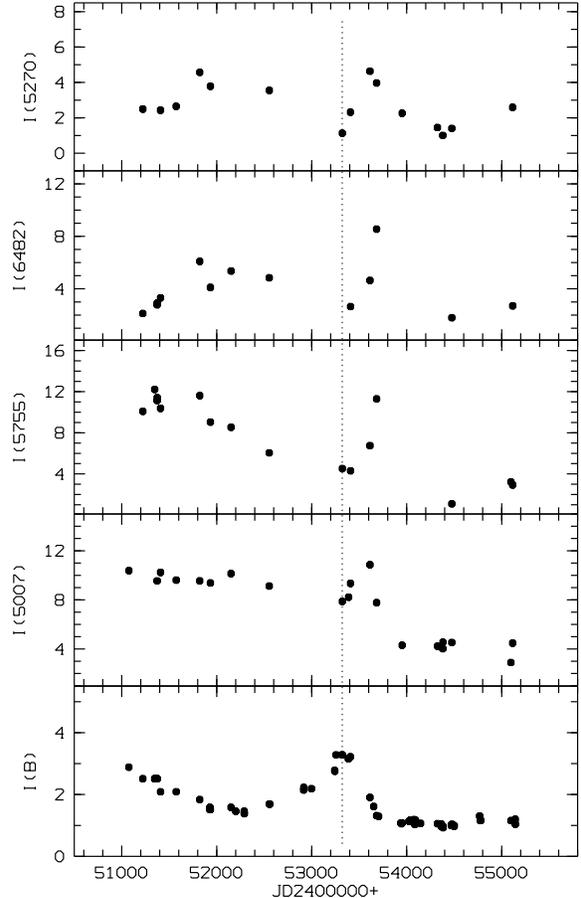}
\captionstyle{normal} \caption{Variation of relative intensities
of forbidden lines [OIII]~$\lambda 5007$, [NII]~$\lambda 5755$,
[FeIII]~$\lambda 5270$,  and permitted NII~$\lambda 6482$ line,
the rest as specified in Fig.\,\ref{9:Fabrika_n_en}. }
\label{10:Fabrika_n_en}
\end{figure}

Figures\,\ref{9:Fabrika_n_en} and \ref{10:Fabrika_n_en} represent
the variations in relative intensity of major lines from 1999 to
2009 with time. In this time interval, our observations allow us
to trace the variations in the spectrum with the variations of
apparent brightness of the star. Relative intensities are obtained
from the equivalent line widths, which were corrected for
brightness variations. We assumed that during the absolute minimum
of $2007-2008$ (with the average brightness  B~$= 18.38^m$,
\mbox{V~$= 18.55^m$}) the star V\,532 is in its true state. All
the equivalent widths are reduced to the state of this minimum
brightness in the V or B-band. To this end, we used a photometric
light curve, and in the spectral observations where there are no
measurements in the V-band, we used the approximation obtained
above,
\small
$$
{\rm
B}-{\rm V} = 0.037 + 0.033({\rm B} - 16.0)^2 - 0.027({\rm B} -
16.0)^3
$$
\normalsize
\noindent (Fig.\,\ref{3:Fabrika_n_en}). The I(B) value
in these figures is the intensity of stellar emission in the
B-band, reduced to \mbox{B~$= 18.38^m$,} i.e., \mbox{I(B~$=
18.38)=1$,} and the relative line intensities in
Figs.\,\ref{9:Fabrika_n_en}, \ref{10:Fabrika_n_en} and
\ref{12:Fabrika_n_en} are their equivalent widths multiplied by
I(B).

Fig.\,\ref{9:Fabrika_n_en} shows the evolution of He\,I lines and the
He\,II~$\lambda 4686$ line. The next figure shows the evolution of
the forbidden lines [OIII]~$\lambda 5007$, [NII]~$\lambda 5755$,
[FeIII]~$\lambda 5270$ and permitted line   NII~$\lambda 6482$. An
estimated position of the center of local brightness maximum of
$2004-2005$ is shown by the vertical \linebreak dotted line.

The most notable feature of the line intensity variations is that
all the lines behave roughly according to their ionization
potentials (or excitation, if these are neutral atoms) and the
temperature of the stellar photosphere. In the local maximum the
temperature significantly decreases, and consequently, the
intensity of He\,I lines increases while the intensity of He\,II
lines drops. Further, when the temperature of the photosphere
significantly increases during the absolute minimum, these lines
vary inversely. The potential of the first ionization  of the
helium atom is 24.6~eV, (the second ionization---54.4 eV), hence,
at the increasing temperature (Figs.\,\ref{9:Fabrika_n_en} and
\ref{12:Fabrika_n_en}) HeI decays into HeII. Ionization potentials
of ions, shown in Fig.\,\ref{10:Fabrika_n_en} amount to 30--55~eV,
but to reach this ionization, 15--35~eV are required. Due to the
rough closeness of their ionization potentials, the behavior of
lines in this figure is similar, although the conditions of
formation of the forbidden and permitted lines, as well as these
three forbidden lines ([NII], [FeIII], [OIII]) between each other
are different. Such a behavior of fundamentally different lines,
demonstrated in Fig.\,\ref{10:Fabrika_n_en}, suggests that these
lines are formed at about the same distance from the stellar
photosphere, and their behavior is determined by the  temperature
variations in the photosphere.

The local maximum of 2004--2005 is fairly sharp, the transition to
the state of absolute minimum is fast, as opposed to the slow
brightness fading in \mbox{$1998-2001$.} In the transition to the
absolute minimum, there is a rapid temperature increase in the
photosphere, the line intensity peak in Fig.\,\ref{10:Fabrika_n_en} is delayed
relative to the continuum intensity peak I(B). There comes a
feeling as if the ``stellar outburst'' (the local maximum of
$2004-2005$) leads to the brightening and excitation of distant
parts of the gas envelope around V\,532, which was ejected in the
earlier stages of activity \cite{Polcaro2010:Fabrika_n_en}. This interpretation
seems to be very controversial, since at the time of local maximum
the temperature of the photosphere did not increase, on the
contrary, it dropped, and quite notably (see below). The values of
``lags'' themselves are very large, from 220 days for [OIII] and
[FeIII] to 350 days for [NII]~$\lambda 5755$ (Fig.\,\ref{10:Fabrika_n_en}). The
physical dimensions of such an envelope would correspond to
$0.2-0.3$~pc. At this distance, for a noticeable additional
ionization of gas a more powerful flash of radiation is needed.

The delay of the peak line intensities in
Fig.\,\ref{10:Fabrika_n_en} is probably due to the rapid
temperature increase in the photosphere of V\,532 after the local
maximum. Line intensities increase when the temperature of the
photosphere becomes optimal for the given ion. Indeed, the
difference in the lag values of [OIII], [FeIII], and [NII] roughly
corresponds (see Fig.\,\ref{12:Fabrika_n_en} below) to different
photosphere temperatures at the times of maxima of these lines.
The He\,I, and He\,II lines in Fig.\,\ref{9:Fabrika_n_en} as well
behave accordingly with the temperature of stellar photosphere at
any given moment of time (despite the fact that
Fig.\,\ref{12:Fabrika_n_en} presents the temperature variations of
V\,532 with time, we would like to emphasize that we describe here
the qualitative behavior of the temperature only, as the
quantitative analysis requires far more detailed calculations).
Consequently, the envelope surrounding the star, in which all
these lines form is compact, it is in fact an extended atmosphere
of the star (the wind). During the local  maximum, He\,II~$\lambda
4686$ line is very weak, but clearly broadened, as discussed above
(Fig.\,\ref{8:Fabrika_n_en}b). In the state of absolute maximum
this line is very bright and has an explicit broad component. For
Fig.\,\ref{9:Fabrika_n_en} we approximated it as a one-component
line, since a two-component profile of He\,II can be reliably
studied only from the spectra with the highest S/N ratio. Below,
we discuss the broad component 4686 and its behavior.

\begin{figure*}[t]
\setcaptionmargin{15mm}
\onelinecaptionsfalse
\includegraphics[scale=0.6,angle=-90,trim=0mm 0mm 0mm 0mm]{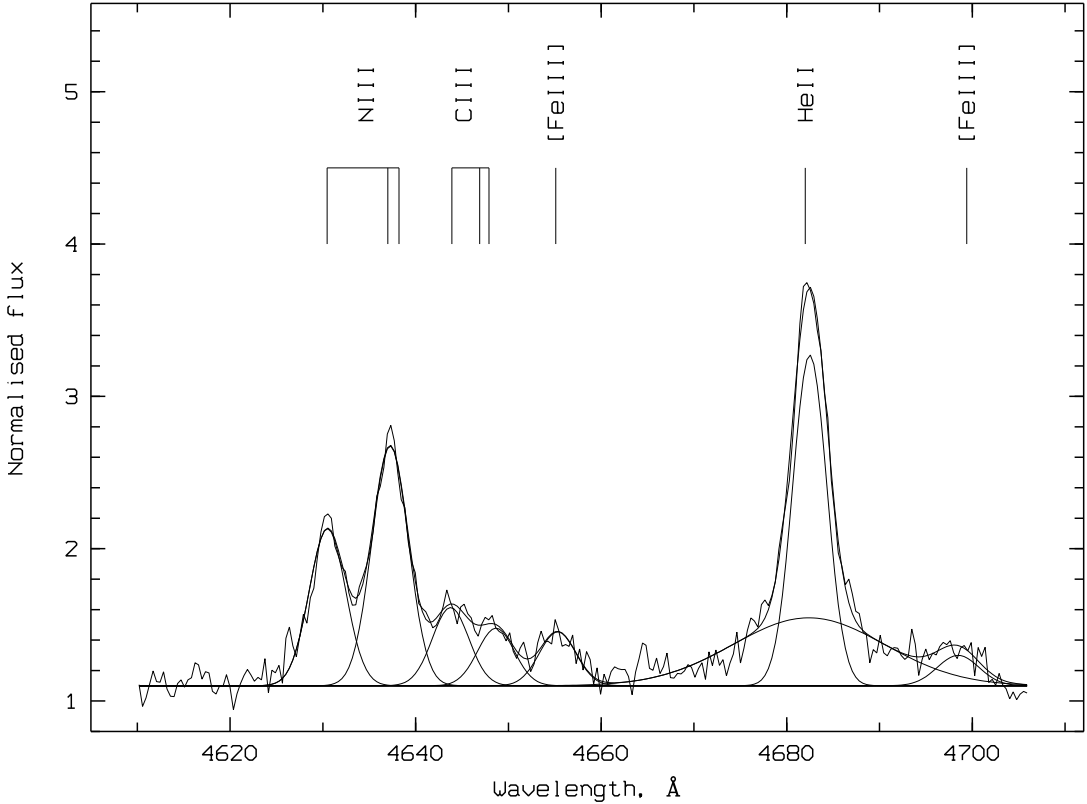}
\captionstyle{normal} \caption{The fragment of the spectrum of
V\,532 (Subaru, 10.08.2007), including a Bowen blend and the
He\,II~$\lambda 4686$ line with an example of a two-component
Gaussian line profile analysis of the latter line.} \label{11:Fabrika_n_en}
\end{figure*}

Figure\,\ref{11:Fabrika_n_en} presents a fragment of the spectrum
of V\,532 (Subaru, 10.08.07, the resolution in this spectrum is
1.1~\AA\,), which includes a Bowen blend and the He\,II~$\lambda
4686$ line with an example of a two-component Gaussian line
profile analysis of the latter line. The same line width was set
for all seven lines, presented in the figure (the narrow component
of He\,II is one of the seven lines), but for the He\,II line we
added a broad component, the position of which coincides with the
position of the narrow component. It was found that ``narrow''
lines have  \mbox{FWHM~$= 4.15~$\AA\,} (taking into account the
spectral resolution) or 270\,km/s, while the broad component of
this spectrum has   \mbox{FWHM~$= 19.5$~\AA\,} or 1250\,km/s. As
noted above, the average width of the broad component of this
He\,II line based on all spectra, in which it was possible to make
the two-component Gaussian analysis, amounts to 1500\,km/s.

The He\,II~$\lambda 5412$ line has an absorption component from
the blue side (a P\,Cyg type profile), however, the \mbox{4686
\AA{}} line did not reveal such a feature. In addition, the 5412
\AA{} line is narrow. Even if we assume that the  5412 \AA{} line
is more narrow due to the absorption component, and try to
construct the original He\,II~$\lambda 5412$ profile, then the
width of such a profile would not be over 10\,\AA\, (less than
550\,km/s).

Above we explain the emergence of a broad component in the
He\,II~$\lambda 4686$ line by the broadening due to the scattering
of photons by electrons. Broad components appear both in the
hydrogen lines, and in the He\,I~$\lambda 5876$ line in such
states of the star, when the temperature of the photosphere is
optimal for the formation of these lines. Which line shows a broad
component is determined by the temperature of dense parts of the
wind in the given state of the star. We explained the absence of a
broad component in the 5412 \AA{} line by a significantly smaller
optical thickness of this line compared to the 4686 \AA{} line,
what is observed in the winds of WN stars
\cite{Hamann1995:Fabrika_n_en}. The narrow profile of the 5412
\AA{} line, as well as the presence of absorption in its blue wing
is fully consistent with this explanation. It means that the 5412
\AA{} line is formed substantially closer to the stellar
photosphere than the \mbox{4686 \AA{}} line. A small width of the
5412 \AA{} line indicates that the wind velocity close to
photosphere of V\,532 is relatively low and a broad component
\mbox{4686 \AA{}} does appear due to the Thomson scattering.

\begin{figure}[t]
\setcaptionmargin{5mm} \onelinecaptionstrue
\includegraphics[scale=0.47,angle=0,bb=70 40 545 655, clip]{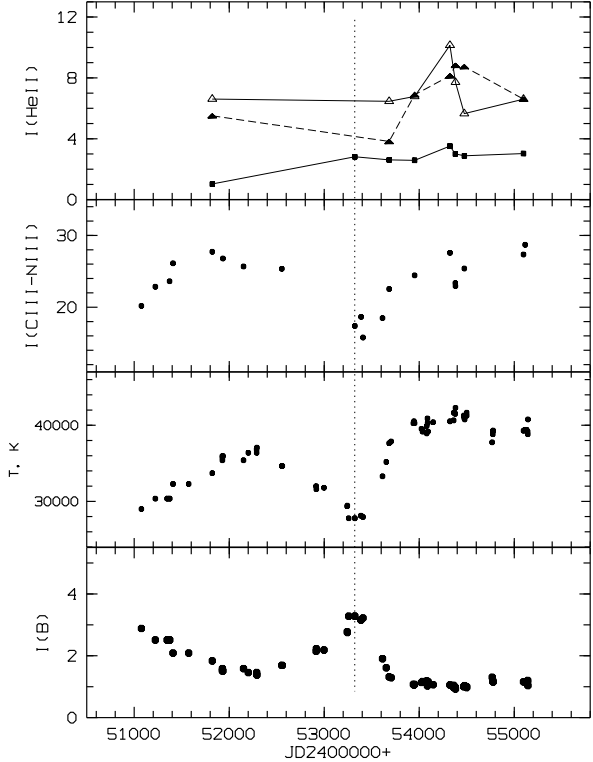}
\captionstyle{normal} \caption{Intensity variation of broad (empty
triangles) and narrow (filled triangles)  components of the
He\,II~$\lambda 4686$ line; intensity variation of the
He\,II~$\lambda 5412$ lines (points) and the Bowen blend with
time. Also shown is the behavior of temperature of the photosphere
in V\,532 (see text for details). The rest as specified in
Fig.\,\ref{9:Fabrika_n_en}.} \label{12:Fabrika_n_en}
\end{figure}

Figure\,\ref{12:Fabrika_n_en} demonstrates intensity variations of
the broad and narrow components of the  He\,II~$\lambda 4686$
line, intensity variation of the He\,II~$\lambda 5412$ line and
the Bowen blend (measured in the wavelength range of
4605--4665~\AA). This figure is similar to
Figs.\,\ref{9:Fabrika_n_en} and \ref{10:Fabrika_n_en}, as before,
the relative intensities I of the lines are derived from their
equivalent widths as I~$=$~EW~$\times$~I(B). We can see that the
5412 \AA{} line does not behave exactly like the 4686~\AA{} line
in general (Fig.\,\ref{9:Fabrika_n_en}). Its intensity is maximal
during the absolute minimum, as well as in line 4686~\AA{}, but in
the local maximum of 2004 its intensity is also high (the
4686~\AA{} line is weak at that time). The broad component of
4686~\AA{} notably increases during the absolute minimum, as it
should if it is formed due to Thomson scattering. The narrow
component of 4686~\AA{} is to a large extent similar to the
behavior of the 5412~\AA{} line, it does not reveal such a strong
maximum at the time when the temperature of the photosphere
dramatically increases. This is also consistent with the idea of
formation of broad component due to the light scattering by
electrons, because then a part of energy in the  4686~\AA{} line
is pumped into its broad component. The behavior of the Bowen
blend CIII/NIII is consistent with the behavior of the photosphere
temperature. The optimum temperature for the formation of this
blend \mbox{($30000-35000$~K)} is notably lower than the optimum
temperature for the formation of the He\,II~$\lambda 4686$  line
(about 50000~K). Therefore, during the absolute minimum the
intensity of the blend CIII/NIII increases not as much as the
intensity of the He\,II line.

The B-band radiation at the photosphere temperature of about
$30000-40000$\,K (the radiation peak is within $700-1000$ \AA)~
falls into the Rayleigh-Jeans part of the spectrum. Hence, we can
use the approximation   I(B)~$\propto T R^2$, where $R$ is the
radius of the stellar photosphere, and $T$ is the temperature of
its photosphere. The variability of S\,Dor-type LBVs (not in giant
eruptions that occurred in $\eta$~Car and P\,Cyg)  fulfills the
condition $T^4 R^2 =$~const in the first approximation. Thus, we
can quite reliably estimate the temperature of an LBV from its
optical brightness, for example, in the B-band, from the ratio $T
\propto$~I(B)$^{-1/3}$, provided that the temperature was once
found independently.

In \cite{Valeev2009:Fabrika_n_en} the temperature of the
photosphere of V\,532 was estimated as $T \sim 35000$~K by fitting
its SED from the optical observations
\cite{Massey2006:Fabrika_n_en}, obtained from autumn 2000 to
autumn 2001, when the star was in the intermediate brightness
state (B~$=$ 17.8).  We plan to determine more accurately the
basic parameters of V\,532 in another paper. Nevertheless, the
temperature variation with time (Fig.\,\ref{12:Fabrika_n_en}),
obtained based on this estimate is consistent with the behavior of
lines of different excitation and ionization potentials,
demonstrated in Figs.\,\ref{9:Fabrika_n_en}, \ref{10:Fabrika_n_en}
and \ref{12:Fabrika_n_en}. The delay of peak intensities of lines,
presented in Fig.\,\ref{10:Fabrika_n_en} is perfectly consistent
with the temperature variation in Fig.\,\ref{12:Fabrika_n_en}. To
the time of maximal intensities of [OIII], [FeIII] and
[NII]~$\lambda 5755$ lines the temperature of the photosphere has
increased from about 30000~K to 38000~K. Further, in the absolute
minimum the temperature of the photosphere has increased even
more, up to about 42000~K and all of these ions had to get to the
next ionization stage. Consequently, the intensities of these
lines have sharply weakened (Fig.\,\ref{10:Fabrika_n_en}).

\subsection{Wind Velocity Variations, Acceleration of the Wind }

In this section we estimate the wind velocity in V\,532, and its
variations depending on the state of the star, i.e. its
brightness. Under the wind velocity we understand the velocity
difference between the absorption line core and the emission peak
$\Delta V_{ae}$, measured from the  P\,Cyg profiles of different
lines. The value $\Delta V_{ae}$ is not an exact wind velocity,
but it is directly related to it.  $\Delta V_{ae}$ traces the
expansion of the atmosphere in the place of the line formation.
The terminal wind velocity is usually somewhat greater than
the escape velocity of the star \mbox{($V_{esc} = \sqrt{2GM/R}$),}
being also directly related to it.

Using simple ratios, described in the previous section, we can
find that $V_{esc} \propto$~I(B)$^{-1/3}$, hence, it depends on
the optical brightness as well as the temperature of the
photosphere. In contrast to the relations $T
\propto$~I(B)$^{-1/3}$ and $R \propto$~I(B)$^{2/3}$, which must be
correct for LBV stars, the dependence $V_{esc}
\propto$~I(B)$^{-1/3}$ may well be inaccurate, as the final wind
velocity depends on a number of conditions in the atmosphere.
Moreover, if we identify $V_{esc}$ with $\Delta V_{ae}$, which can
be directly measured, i.e. if we use a relation $\Delta V_{ae}
\propto$~I(B)$^{-1/3}$ we can make a significant quantitative
error. But qualitatively, with an increase in the size of the
stellar photosphere (which depends on the optical brightness as $R
\propto$~I(B)$^{2/3}$) the wind velocity should decline.

\begin{figure*}[ht]
\setcaptionmargin{9mm} \onelinecaptionsfalse
\includegraphics[scale=0.6,angle=-90,trim=0mm 0mm 0mm 0mm]{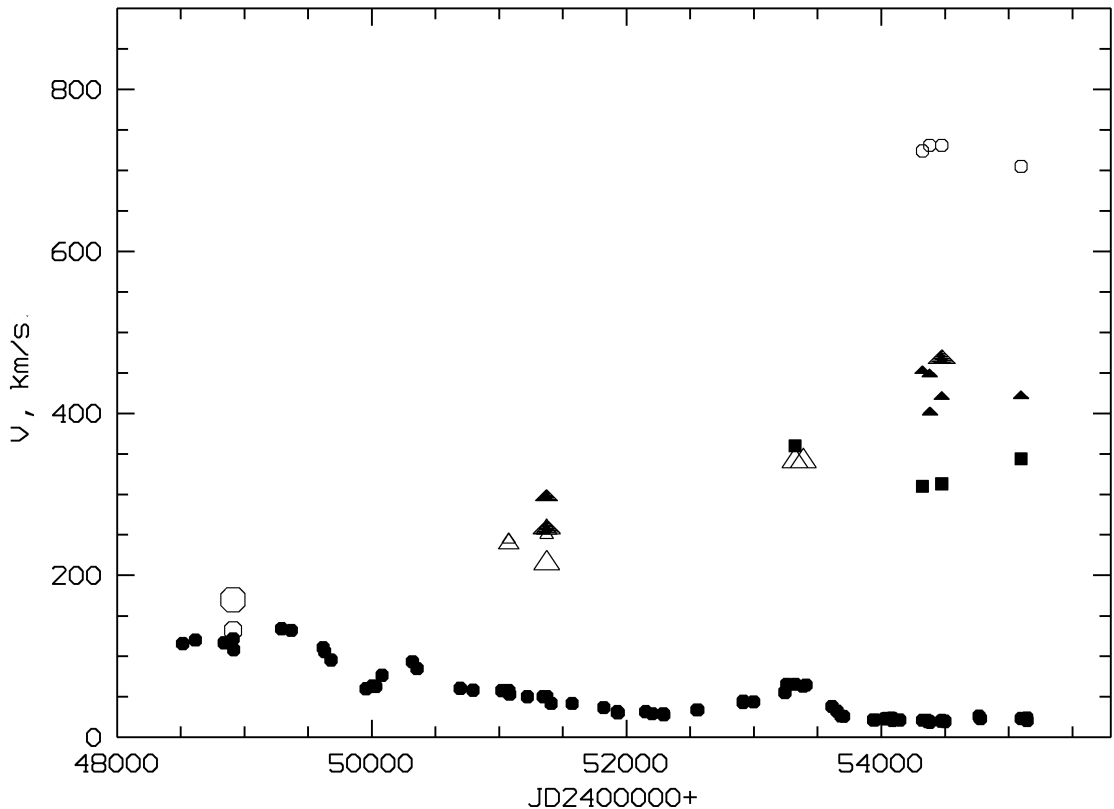}
\captionstyle{normal} \vspace{5mm}\caption{The variation of wind
velocity of V\,532, measured as $\Delta V_{ae}$ with time. The
filled circles show the I(B) dependence, for comfortable viewing
the value of I(B) is multiplied by a factor of 20. The velocity is
measured from all the spectra we have, and in all lines, revealing
P\,Cyg type profiles. Large empty circles mark the H$\alpha$ line,
medium-sized circles---H$\beta$ line, small circles---H$\delta$
line, large empty triangles---He\,I~$\lambda 6678$, medium empty
triangles---He\,I~$\lambda 5015$, small empty
triangles---He\,I~$\lambda 4922$, large filled
triangles---He\,I~$\lambda 5876$, medium filled
triangles---He\,I~$\lambda 4713$, small filled
\mbox{triangles---He\,I~$\lambda 4471$,} and filled squares mark
the He\,II~$\lambda 5412$ line. Different lines reveal P\,Cyg type
profiles depending on whether their excitation/ionization
potential corresponds to the current value of the photosphere
temperature (see
Figs.~\ref{9:Fabrika_n_en},\ref{10:Fabrika_n_en},\ref{12:Fabrika_n_en}).
During the absolute brightness minimum only the He\,I triplet
lines possess P\,Cyg type profiles. } \label{13:Fabrika_n_en}
\end{figure*}


Figure\,\ref{13:Fabrika_n_en} demonstrates wind velocity
variations of V\,532 with time, measured as $\Delta V_{ae}$. The
velocity is measured from all the spectra we have, and in all the
lines revealing P\,Cyg type profiles. If the spectra contain many
lines of one type (e.g., the lines of hydrogen or He\,I) with a
P\,Cyg profile, then, in order not to overload the figure, we show
the results of measurements only for those lines, in which the
absorption core can be most reliably measured. The same figure
demonstrates the variation of I(B) with time, for convenience the
value of I(B) is multiplied by a factor of 20. We can see that in
the state of absolute maximum the wind velocity is the lowest. At
this time the P\,Cyg type profile is perceivable only in hydrogen
lines. The rate of expansion of the atmosphere is higher in the
line of H$\alpha$ than in H$\beta$, which is natural, as the first
line is formed in higher layers of the atmosphere. In the state of
intermediate brightness and during the local maximum of
\mbox{$2004-2005$,} P\,Cyg profiles were revealed only by He\,I
lines, both by singlets and triplets.

In the state of absolute minimum P\,Cyg profile was observed in
He\,I, He\,II~$\lambda 5412$ lines and in the hydrogen H$\delta$
line. It is important that in this state P\,Cyg profiles were
revealed only by the triplet lines of He\,I, which is due to the
difference from the singlet lines in their formation. We can see
(Fig.\,\ref{13:Fabrika_n_en}) that in the state of absolute
minimum a clear correlation is observed between the excitation
potential and the velocity in $\Delta V_{ae}$, the higher the
potential, the lower the velocity of wind.

\begin{figure*}[ht]
\setcaptionmargin{8mm} \onelinecaptionsfalse
\includegraphics[scale=0.6,angle=-90,trim=0mm 0mm 0mm 0mm]{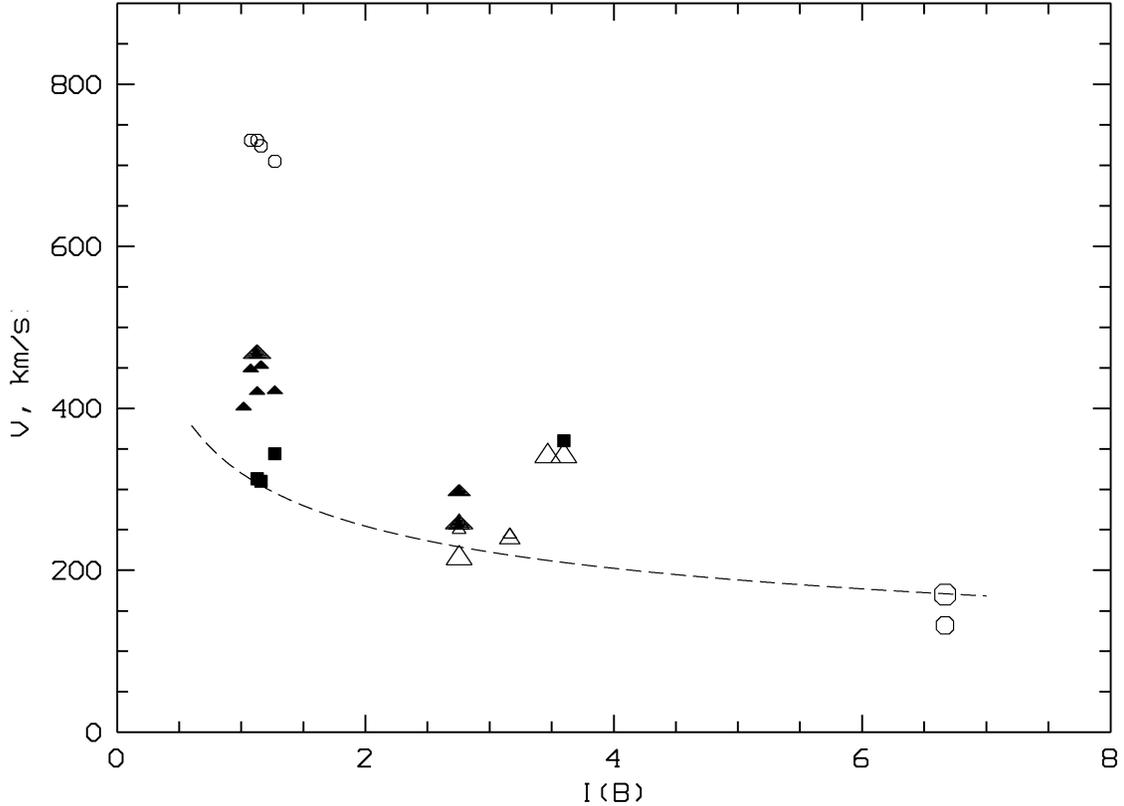}
\captionstyle{normal} \caption{The dependence of the atmosphere
expansion velocity on the relative optical brightness. The
markings are the same as in the previous figure. In the state of
absolute minimum we can observe the kinematic profile of the
atmosphere (the wind), the wind temperature decreases with
increasing distance from the photosphere, the lower the gas
temperature in the wind, the faster the expansion of the
atmosphere. The wind velocity depends on a number of factors, and
it can be only nominally expressed by a simple dependence $\Delta
V_{ae} \propto$~I(B)$^{-1/3}$, which is shown by a dashed line.
The value of  $\Delta V_{ae}$ determines the wind velocity in the
region of line formation.} \label{14:Fabrika_n_en}
\end{figure*}

Figure\,\ref{14:Fabrika_n_en}  demonstrates the dependence of the
wind velocity on the relative optical brightness. We can see that
with decreasing brightness I(B), that is, with declining size of
the photosphere the wind expansion velocity is clearly increasing.
As noted above, the radius of the LBV star is determined quite
well by the dependence $R \propto$~I(B)$^{2/3}$, but the wind
velocity  can only very conditionally  be expressed by the
relation $\Delta V_{ae} \propto$~I(B)$^{-1/3}$, the more so since
the value $\Delta V_{ae}$ determines only the velocity of
expansion in the region of formation of this line. Nevertheless,
the expansion velocities, measured from the H$\alpha$ line in the
state of absolute maximum, He\,I  lines in the intermediate state,
and He\,II~$\lambda 5412$ line in the state of absolute minimum
are consistent with the dependence $\Delta V_{ae}
\propto$~I(B)$^{-1/3}$. In each of the these states, these lines
are formed in the deeper parts of the wind, and perhaps for this
reason this simple relationship holds (however, H$\alpha$ line in
the state of maximum brightness is formed higher than H$\beta$
line).

In the state of absolute minimum we can see the kinematic profile
of the atmosphere (Fig.\,\ref{13:Fabrika_n_en} and
\ref{14:Fabrika_n_en}): the lowest wind velocity ($330-350$ km/s)
is observed in the He\,II~$\lambda 5412$ line, intermediate
velocity in this state ($400-470$~km/s) is observed in the triplet
He\,I lines, and the maximum velocity ($700-730$ km/s) is observed
in H$\delta$ line. The interpretation of this dependence is clear,
the wind velocity increases outwards according to the acceleration
of wind. In addition, it clearly shows that the temperature of the
wind in the extended atmosphere of V\,532 declines with distance
from the star. The gas temperature in the wind may be determined
by many factors. They define the energy balance between the
heating through radiation (heating by shock waves is also
possible, if such exist in the winds of LBV) and the gas cooling.
The cooling of gas occurs both through radiation, and due to
expansion. One way or another, we can confidently conclude that we
have found explicit evidence of acceleration and cooling of the
wind of V\,532  based on the spectra of V\,532 in absolute
minimum.

\section{CONCLUSION}

This paper presents the results of the most complete spectroscopy
of an LBV star V\,532 in M\,33. The spectra cover both the
absolute maximum occurring between \mbox{$1992-1994$} (the
high/cold state), and the absolute minimum of $2007-2008$ (the
low/hot state). The brightness difference between these extremes
amounts to $\Delta {\rm B} \approx 2.3^m$. In addition to the
extreme brightness states of  V\,532, spectral observations also
cover the local minimum and maximum of $1998-2006$. The
variability of the spectrum of V\,532 completely corresponds to
the variations in temperature of its photosphere depending on the
visual brightness ($T \propto$~I(B)$^{-1/3}$) ~in the B-band.

Following the estimation of the photosphere temperature of V\,532
from~\cite{Valeev2009:Fabrika_n_en}, $T \sim 35000$~K, calculated
from the data obtained when the star was in its intermediate state
(${\rm B} = 17.8$),  in the absolute minimum its temperature
reached \mbox{$T \sim 42000$~K,} while in the absolute maximum the
photosphere cooled down to $T \sim 22000$~K. Despite the fact that
these estimates  of the photosphere temperature are quite rough (a
special simulation of the  V\,532 spectrum is needed), the
appearance, variation and disappearance of major lines in the
stellar spectrum is fully consistent with their ionization
potentials (excitation in the case of neutral atoms) and current
temperature of the photosphere. Even the forbidden lines of
[OIII], [FeIII], [ArIII] and [NII]~$\lambda$5755 behave
accordingly with the permitted lines and follow the same
dependence on the temperature of the photosphere. We hence
conclude that the envelope around the star, in which all these
lines are formed is relatively compact, it is an extended
atmosphere of the star (its wind). In contrast to the line
[NII]~$\lambda$5755, the line [NII]~$\lambda 6583$  belongs to the
extended nebula sized tens of pc \cite{Fabrika2005:Fabrika_n_en},
which, however, is captured by the spectrograph slit and can not
be removed.

We have traced the evolution of broad components of the brightest
lines. In the spectrum of V\,532 they appear in the strongest
lines of hydrogen, HeI and HeII~$\lambda$4686, their width ranging
from 1100 to 1800~km/s. Broad components also behave in parallel
with
the temperature of the photosphere. In the high state a broad
component appears in hydrogen lines, when the temperature of the
star increases, the broad wings disappear. In the intermediate
state  a broad component appears in the He\,I~$\lambda$5876 line,
this component is not observed in hotter or colder states. In the
low state, broad wings appear in the HeII~$\lambda$4686 line,
however, they are not observed in the line of the same ion
HeII~$\lambda$5412. It follows from this that broad components
appear in the lines with maximal optical depth, and only in those
states/lines, when the temperature of the photosphere is optimal
for the excitation of the given transition. Consequently, broad
components are in no way connected with stellar rotation or with
the Doppler broadening of spectral lines in the wind, since the
bright narrow components are present in the same line profiles.
The appearance of broad line wings is related to the broadening
due to the Thomson scattering in the most dense and closest to the
photosphere parts of the wind.

The HeII~$\lambda$5412 emission is relatively narrow. In contrast
to the 4686~\AA{} line, its profile reveals an absorption
component, shifted towards the blue region. We explain the absence
of a broad component in the 5412~\AA{} line by a significantly
lower optical depth of this line, compared with the 4686~\AA{}
line, as it is known in the winds of WN stars
\cite{Hamann1995:Fabrika_n_en}. A narrow profile of the 5412~\AA{}
line, and the presence of absorption in its blue wing is fully
consistent with this explanation, i.e. the 5412~\AA{} line is
formed much closer to the photosphere of the star than the
4686~\AA{} line. A small width of the 5412~\AA{} line indicates
that the wind velocity close to the photosphere is relatively low,
and the broad component of 4686~\AA{} does indeed appear due to
the Thomson scattering.

Radial velocity of V\,532 is $-184 \pm3$~km/s, it is constant with
an accuracy of a few km/s. This value was found from the two best
spectra, obtained with the time difference of 15 years. The wind
velocity of V\,532, measured as $\Delta V_{ae}$ varies with time.
The velocity is measured from all the spectra we have, and in all
the lines, revealing P\,Cyg type profiles. The wind velocity
clearly depends on the radius of the stellar photosphere  (which
is determined sufficiently well by the relationship $R
\propto$~I(B)$^{2/3}$), i.e. on the value of visual brightness of
the star. With decreasing brightness I(B), or with declining size
of the photosphere the wind velocity increases. This is discovered
based on different lines, in so far as when the temperature of the
photosphere changes, different lines reveal P\,Cyg-type profiles.

In the absolute minimum we managed to find the
kinematic profile of the atmosphere of V\,532. The lowest wind
velocity ($330-350$~km/s) is observed in the He\,II~$\lambda 5412$
line, intermediate velocity \mbox{($400-470$~km/s)} is observed in
the triplet He\,I lines, while the maximum wind velocity
\mbox{($700-730$~km/s)} is observed in the H$\delta$ line. This
implies that the wind velocity increases outwards,
and that the wind temperature in the extended atmosphere of V\,532
decreases with distance from the star.

Using  quantitative spectral criteria, introduced \mbox{in
\cite{Crowther1997:Fabrika_n_en},} it was found in \cite{Fabrika2005:Fabrika_n_en} that in
the intermediate brightness state  of $1998-2001$, the spectral
type of the star corresponds to WN10-11. When the star was in the
low/hot state, \mbox{Polcaro et al. \cite{Polcaro2003:Fabrika_n_en}}, using the
same criteria \cite{Crowther1997:Fabrika_n_en} found that the spectral type of
V\,532 was close to WN8-9. It means that the spectral class of
V\,532 varies between WN11 and WN8-9 depending on visual
brightness. We confirm these results based on a more
representative set of spectra. Using the criteria of spectral
classification of WN stars from \cite{SmithMoffat1996:Fabrika_n_en} we have
identified the spectral class of V\,532 in its low state as
WN8.5h, based on several criteria.

Using the Crowther and Smith diagrams
\cite{Crowther1997:Fabrika_n_en} we have traced the evolution of
V\,532 along with the evolution of two other LBV stars, AG\,Car
and the WN component of the massive binary HD\,5980. AG\,Car has
over the past few years shown the \mbox{${\rm LBV} \leftrightarrow
{\rm WNL}$-type} transition. HD\,5980 has revealed the reverse
 transition, more precisely, an LBV episode
\mbox{${\rm WN3} \to {\rm WN11(LBV)} \to {\rm WN4/5}$} in autumn
1994. Hence, only three objects are known  by now that can be
studied in detail in their \mbox{${\rm LBV} \leftrightarrow {\rm
WNL}$-type} transitions, namely, V\,532, AG\,Car and HD\,5980.

We found that during the visual minima all the three stars
perfectly fit into the sequence of WNL stars
\cite{Crowther1997:Fabrika_n_en}. However, when the visual
brightness increases, all the three stars, AG\,Car, V\,532 and
HD\,5980 form a separate sequence, not consistent with the WNL
sequence \cite{Crowther1997:Fabrika_n_en}, and hence they depart
beyond its limits. Meanwhile, V\,532, both in the local maximum,
and with increasing visual brightness falls into one and the same
new sequence.

The departure from the WNL sequence with increasing brightness is
related to the HeII~$\lambda$4686  line broadening, when its width
does not correspond to its equivalent width, which is required to
remain within the  WNL sequence. Perhaps this is due to the wind
which is denser, than those, observed in the WNL stars, i.e., due
to the line broadening via the Thomson scattering. It is possible
that what we see is a new property of LBV stars, when in the
high/cold state they do not correspond to the latest {\it bona
fide} WNL stars. We presently know only three examples of such
transitions. Obviously, to make more reliable conclusions one
requires additional observations and new objects.


\begin{acknowledgments}
The authors thank T.~Szeifert for the kindly presented spectra,
A.~Knyazev for spectroscopic observations performed with the CAFOS
on the 2.2-m  Calar-Alto telescope and the reduction of obtained
spectra, E.A.~Barsukova for her assistance with photometric
observations and spectra processing, and E.L.~Chentsov for useful
discussions. The work was supported by the  RFBR grants nos.
\mbox{09-02-00163,} \mbox{10-02-00463,} the grant Leading Research
Schools of Russia no.~5473.2010.2., and by the federal grant
Scientific and Teaching Staff of Innovative Russia $2009-2013$,
P1244.

\end{acknowledgments}

\end{document}